\documentclass[10pt]{iopart}
\usepackage{epsfig}
\usepackage{iopams}
\usepackage{graphicx}

\renewcommand{\epsfig}[1]{}

\newcommand{\be}{\begin{equation}}
\newcommand{\ee}{\end{equation}}
\newcommand{\bea}{\begin{eqnarray}}
\newcommand{\eea}{\end{eqnarray}}

\newcommand{\eq}[1]{(\ref{#1})}

\catcode`@=11 
\renewcommand\tableofcontents{%
  \section*{\contentsname}%
  \@starttoc{toc}%
}
\catcode`@=12

\begin{document}

\topical[Single-molecule experiments] {Single-molecule experiments in
biological physics: methods and applications.}

\author{F. Ritort}
\address{ Departament de F\'{\i}sica Fonamental, Facultat de
  F\'{\i}sica, Universitat de Barcelona, Diagonal 647, 08028 Barcelona,
  Spain
}

\ead{\mailto{ritort@ffn.ub.es}
}

\begin{abstract}
I review single-molecule experiments (SME) in biological physics.
Recent technological developments have provided the tools to
design and build scientific instruments of high enough sensitivity and
precision to manipulate and visualize individual molecules and measure
microscopic forces. Using SME it is possible to: manipulate molecules
one at a time and measure distributions describing molecular
properties; characterize the kinetics of biomolecular reactions and;
detect molecular intermediates. SME provide the additional information
about thermodynamics and kinetics of biomolecular processes. This
complements information obtained in traditional bulk assays. In
SME it is also possible to measure small energies and detect large
Brownian deviations in biomolecular reactions, thereby offering new
methods and systems to scrutinize the basic foundations of statistical
mechanics. This review is written at a very introductory level
emphasizing the importance of SME to scientists interested in knowing
the common playground of ideas and the interdisciplinary topics
accessible by these techniques.

The review discusses SME from an experimental perspective, first
exposing the most common experimental methodologies and later presenting
various molecular systems where such techniques have been applied. I
briefly discuss experimental techniques such as atomic-force microscopy
(AFM), laser optical tweezers (LOT), magnetic tweezers (MT), biomembrane
force probe (BFP) and single-molecule fluorescence (SMF). I then present
several applications of SME to the study of nucleic acids (DNA, RNA and DNA condensation),
proteins (protein-protein interactions, protein
folding and molecular motors). Finally, I discuss applications of SME to
the study of the nonequilibrium thermodynamics of small systems and the
experimental verification of fluctuation theorems. I conclude with a
discussion of open questions and future perspectives.
\end{abstract}


\maketitle

\tableofcontents

\section{Introduction}
\label{intro}

The study of single molecules has become a major theme of research in
modern biophysics. Specialized journals are devoted to this emerging
field and every year more researchers become attracted to it. The
history of single-molecule experiments is related to that of
single-molecule imaging and has its roots in the invention of optical
tweezers and the scanning tunneling microscope. The possibility of
manipulating individual entities has always attracted the
scientist. Since the experimental discovery of the nucleus of the atom
by Rutherford it has become a major goal in physics to search for the
ultimate constituents of matter. A similar trend is followed in modern
biology. There the main goal has been to characterize and understand the
function of all constituent parts of the living organisms and,
ultimately, the chemistry of life. Despite the fact that biology and
physics are very different sciences (most students in biology rarely
feel attracted by physics and vice-versa) the true fact is that these
two sciences are to become much closer than ever in the new starting
century. Although there have been many notable contributions to
molecular biology by physicists (like Max Delbruck, Francis Crick, the
Braggs just to cite a few names) the two sciences have diverged over the
past thirty years. Molecular biology has developed very specific
experimental methods, most of them borrowed from biochemistry. These
methods are largely unknown to physicists (PCR amplification, gene
cloning,..). The current tendency is for this temporary gap to
progressively narrow. More physicists will learn about the subtleties of
biological matter and therefore become acquainted with some of the
techniques and methods used by biologists. The marriage between physics and biology may still take
years, chemistry becoming the privileged witness of the union.

What is the benefit of such a marriage? On the one hand, physicists are
becoming steadily interested in the properties of biological
matter. The kinetics of molecular motors, the folding of biomolecules,
the viscoelastic and rheological properties of the cell, the transport
of matter through pores or channels, the physical properties of
membranes and the structure of biological networks are just a few
examples of subjects akin to the expertise and interests of the
physicist. On the other hand, biologists are interested in the
physical techniques and methods available from physics. Physics is a
quantitative science while biology has been traditionally mostly
descriptive \cite{Alberts98}. It is not surprising that the discovery of the double
helix was made possible thanks to X-ray diffraction, an experimental
technique discovered and used by physicists to determine atomic
structures in crystals. More important, the biologist is steadily
aware of the great complexity of biological matter, the large variety
of biological forms and the relevance of trying to unify such
knowledge. Physical abstraction can be important to
single out common themes and variations throughout this vast phenomenology. In
addition, current experimental methods applied to biological systems
are providing a huge amount of data that must be quantitatively
analyzed by using sophisticated methods. The physicist
can help a lot in this task.

It is fair to say that single-molecule experiments will contribute to
bridge the gap between physics and biology. Single-molecule experiments
(hereafter referred as SME) provide a new tool in physical biochemistry
that allows to explore biochemical processes at an unprecedented
level. They offer a quantitative description of biological processes
reminiscent of physicists approach. SME are made possible thanks in
part to the advent of nanotechnologies. These, combined with microscale
manufacturing techniques, provide the technology required to
design and build scientific instruments of enough sensitivity and
precision to manipulate individual molecules and measure microscopic
forces, thereby allowing experimentalists to investigate various
physical and biological processes. Nowadays, the most widespread and
commercially available single-molecule technique in biophysics labs is
the atomic force microscope (AFM). This technique allows one to take
images of individual molecules adsorbed into surfaces. At the same time
with the AFM it is possible to grab molecules one at a time by attaching
one end of the molecule to the AFM tip, the other being immobilized on
the surface. By moving the tip relative to the substrate it is then
possible to pull the molecule away from the surface and exert mechanical force. The value of the breakage force, the distance that the
tip has to be retracted before the contact breaks and the dependence of
these numbers on the speed of the moving tip are important information
about the mechanical strength and location of the probed molecular bond. By pulling apart many
molecules one at a time it is possible to quantitatively characterize
the breakage process, thereby giving precious information about
molecular interactions.

There are several excellent reviews in SME, applications and
methodologies. Most of them fall into two categories. Either they are
very introductory and cover generic topics on single-molecule
manipulation or they are more specialized and specifically devoted to
discuss particular topics.  It could not be otherwise. The field of
single molecules is developing very fast and at the same time
diversifying into many different areas. Therefore it is very difficult
to cover all the subjects in a review. SME deal with aspects related
to instrumentation, their application to study many different systems
(belonging to physics, chemistry and biology), theoretical modeling
and numerical simulations. Only in the area of optical tweezers, there
is a complete resource letter available with useful references until
the year 2003 (which includes nearly 400 references)
\cite{LanBlo03}. This number of references is steadily growing every
year.

The present review attempts to partially fill this gap by presenting an
 overview of various topics from an experimental perspective, first
exposing the most important experimental methodologies and later
reviewing various molecular systems where such techniques have been
applied. I also include a brief section describing the applications of
SME to investigate thermodynamics at the molecular level.  This review
covers SME applied to biomolecules, it therefore does not touch upon
applications of single-molecule techniques to other interesting subjects
such as living cells or non-biologically inspired problems. It briefly
discusses theoretical modeling and numerical simulations. When
appropriate a few references about theoretical models and simulations are listed for
those readers who want to delve deeper into the subject. The selection of
topics has been naturally biased by my own taste and expertise. Although
I have tried to cover the most relevant existing literature it is
unavoidable that some important work and papers have been unduly
omitted. I apologize in advance to these colleagues whose work may have
been overlooked.

Very introductory reviews to SME can be found in
\cite{BaiWanXieWol99,FisObeCarMarFer99,StrAllCroBen01,BusMacWui00}. There
are also more focused reviews on the mechanical properties of
biomolecules
\cite{BusSmiLipSmi00,Bao02,CocMarMon02,StrDesChaDekAllBenCro03,BrySmiBus03}, the
elastic properties of proteins
\cite{WanForJin01,CarObeFisMarLiFer00,Car05}, mechanochemistry \cite{BusChemForIzh04}, single-molecule
fluorescence \cite{Weiss99,Ha01,Haran03} and instrumentation
\cite{NeuBlo04,Williams}. Whole journal issues devoted to review SME,
some of which have been published as books \cite{LeuZla01}, can
be found in \cite{science99} and proceedings of biophysics conferences
often include a section on SME \cite{Houches02,Houches04}. We must also
mention web pages with detailed information about specific
single-molecule techniques (for example, laser tweezers \cite{lot}) or
excellent review journals \cite{cosbjournal}. Finally, reviews
about the usefulness of SME to investigate the thermodynamics of
molecular systems can be found in \cite{Ritort03,BusLipRit05,Ritort06}.

\section{Why single molecules?}
\label{why}

SME are central to biological physics research \cite{FraWolAus99}.
These offer a complementary yet different perspective to understand
molecular processes. What are the advantages of SME compared to
traditional bulk assays?  The main difference between single molecule
and traditional biochemistry methods lies in the kind of average done when measuring
the properties of the system. SME allow experimentalists to access
biomolecular processes by following individual molecules. Using SME it
is possible to measure distributions describing certain molecular
properties, characterize the kinetics of biomolecular reactions and
observe possible intermediates. SME provide additional information
about thermodynamics and kinetics that is sometimes difficult to obtain
in bulk experiments.  All this is complemented by powerful visualization
methods (which allow to capture images and produce movies) that greatly
help the scientist in the interpretation and understanding of the
experiments.

 To better understand the advantages of SME let us consider the example
of protein folding. A protein in water solution can exist in two
possible conformations (folded and unfolded). In one conformation the
protein is folded into its native state forming a compact globular
structure. Roughly speaking, the hydrophobic core is buried inside the globule and
stabilized by specific amino acid-amino acid interactions whereas the
hydrophilic amino acids are exposed to the outside on the surface of the
globule. In the other conformation the protein is denatured or unfolded
forming a random coil. At room temperature (e.g. 25$^\circ$C or 298$^\circ$K) the
protein is in the native state as this state has a free energy that is
lower than that of the random coil. Upon heating (or increasing the
concentration of denaturants such as urea), the protein can denature and
change conformation from the native to the unfolded state. Most proteins
typically denature at temperatures in the range 50$^\circ$C-80$^\circ$C, each protein
being characterized (under given solvent conditions of salinity and pH)
by a melting temperature, $T_m$, where the protein denatures. The full
characterization of this transition is possible by using calorimetry
bulk measurements where the protein is purified inside a test tube. The
enthalpy curve obtained in such measurements shows a jump in the
enthalpy and a latent heat (similar to that observed in water at its
boiling point). This is the characteristic signature of a first-order
phase transition separating two conformations. The same conclusion is
reached and the same melting temperature found by carrying out other bulk
measurements (such as UV absorbance).

What additional information can be obtained with single-molecule
techniques? It is a well known fact that during the folding process some
proteins transiently visit an intermediate state, the {\em molten
globule} state, characterized by a short lifetime. In such state
proteins form a globular structure where, roughly speaking, the
hydrophobic and hydrophilic parts are separated between the core and the
surface of the globule but specific contacts between amino acids are not
yet fully formed.  Due to its short lifetime, the fraction of molecules
that are in the molten globule state inside the test tube can be small
enough to go unobserved in calorimetry measurements. The tiny signal
they produce can be masked by that of the overwhelming number of
correctly folded or totally unfolded molecules. In SME one can follow
one protein at a time, therefore it is possible to separate molecules
into three families: the native (N), the intermediate molten globule (I)
and the unfolded (U).

\begin{figure}
  \centering \includegraphics[scale=.55,angle=-90]{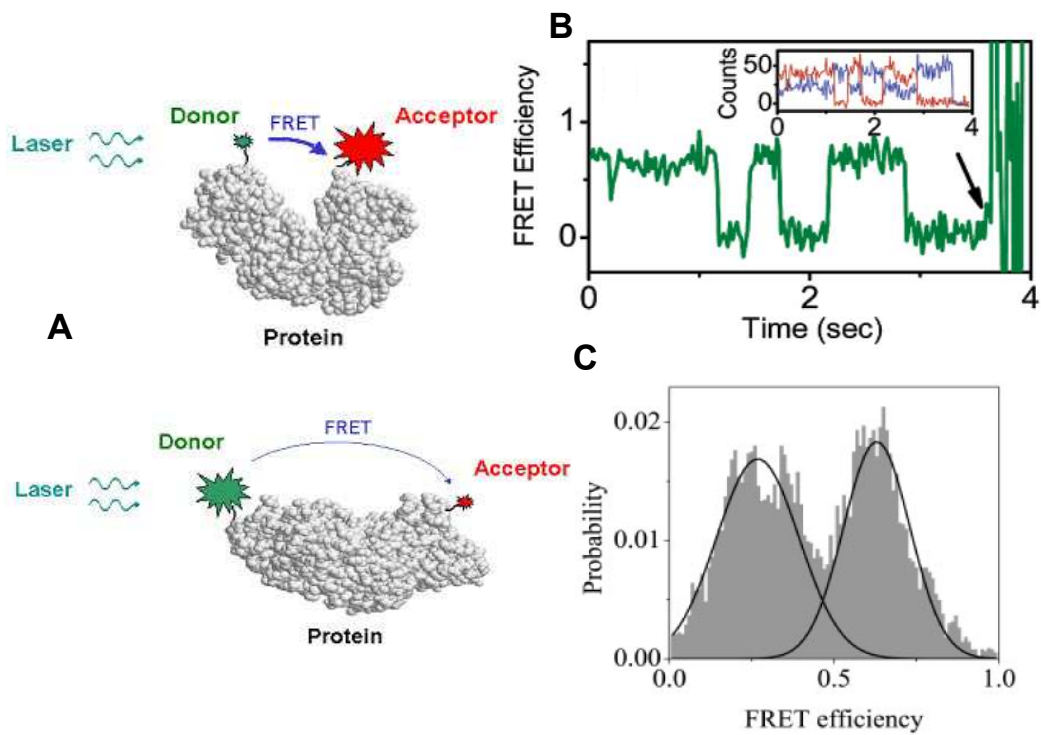}
  \caption{Single-molecule detection using FRET. (A) A conformational
  change in a protein changes the relative intensities of light emitted
  by the donor and the acceptor. (B) Typical FRET efficiency
  trajectories and donor/acceptor intensities $I_D,I_A$ (Inset) for
  individual proteins trapped in vesicles. These show multiple
  folding/unfolding events close to the midpoint folding transition. The
  arrow indicates when photobleaching occurs. (C) Probability
  distributions of FRET efficiencies showing the existence of two
  families of proteins (folded and unfolded). The average FRET
  efficiency is around 0.6 and agrees well with values obtained in bulk
  assays. Figures taken from \cite{RhoCohSchHar04,RhoGusHar03}.}
  \label{fig1}
\end{figure}

By using single-molecule fluorescence it is possible to attach
fluorescent molecules to proteins, detect them by focusing light into
a tiny spot, and watch proteins go through that spot for a short
interval of time. Fluorescent proteins are chemically synthesized by
attaching fluorescent dyes into specific residues of the amino acid
chains of the protein (see Figure~\ref{fig1}). In single-molecule
fluorescence resonance energy transfer (FRET) techniques, two
different color dyes (e.g. {\em green} and {\em red}) can be
positioned at specific locations of the protein. These locations stay
close each other when the protein is in its native state, only
slightly close in the intermediate state and far away in the unfolded
state. Upon radiation of light with the appropriate frequency the {\em
green} dye (the donor) absorbs the radiation. A fraction of that
intensity of light is emitted to the observer ($I_D$). The rest of
intensity is emitted by the {\em red} dye (the acceptor) through a
non-radiative resonance energy transfer mechanism between donor and
acceptor (the condition being that the emission spectrum of the donor
overlaps with the absorption spectrum of the acceptor). The light
emitted by the acceptor has always lower frequency than that emitted
by donor and part of the energy transferred to the acceptor is lost to
the environment in the form of heat. The amount of non-radiative
energy transfer between the {\em green} and the {\em red} dye depends
on the distance between the dyes and defines the FRET efficiency,
$E=I_A/(I_A+\eta I_D)$ where $\eta$ is a correction factor that
depends on the quantum yields of donor and acceptor. For $E=1$ all
light absorbed by the donor is transferred to the acceptor whereas for
$E=0$ the light is emitted only by the donor. The intensity of light
emitted by the donor and the acceptor that is detected by the observer
changes as the distance between the two dyes changes. However the
total amount of light emitted by the donor and the acceptor is
constant, until photobleaching occurs. Therefore the intermittent
exchange between the amount of light emitted at both wavelengths is an
indirect measure of the distance between the dyes, i.e. of the
different conformations of the protein (providing a spectroscopic ruler). 

Coming back to our previous example, measurements taken over many
proteins might show a three-modal distribution indicating three possible
conformations of the molecules (N,I,U). The fraction of molecules
observed in each of these three states (i.e. the statistical weight of
each {\em mode}) would be a function of the temperature and/or
denaturant concentration. Below/Above the melting temperature nearly all
molecules are folded/unfolded. In the vicinity of the melting
transition, intermediate conformations would be observable together with
some native and unfolded conformations as well.  Single-molecule
techniques offer also the possibility to measure kinetic or
time-dependent processes by following the trajectories of individual
molecules.  The signal would execute transitions as the conformation of
the protein changes, providing direct evidence of the existence of an
intermediate state as well as information about the kinetic rates
between the three states. Sometimes the protein would execute
transitions from the unfolded to the itermediate state ($U\to I$) to
later jump to the native state ($I\to N$). In this case the molten
globule is an intermediate {\em on-pathway} to the native state. In
another scenario the protein directly folds into the native state
without first collapsing into the intermediate ($U\to N$). In this case
the molten globule would be an intermediate state {\em off-pathway} to
the native state. The statistics of the residence times of the protein
in each state provides extremely valuable information about the folding
process. All this information is usually masked in bulk measurements
where results are averaged over molecules and time. SME complement
standard spectroscopy and microscopy techniques in molecular biology and
biochemistry and therefore have to be viewed as a new source of valuable
information to interpret biomolecular processes.

\section{From physics to biology and back}
\label{physbioback} 

It is fair to say that single-molecule techniques have been largely
developed by physicists.  These new kind of measurements are providing
lots of quantitative information about molecular processes that have to
be analyzed using statistical methods. This is very attractive to the
physicist who can propose experiments to test new theories or to
investigate theories and analyze models to interpret the experimental
results. This statement can be illustrated with the following
example. It is worth mentioning that one among the first
single-molecule pulling experiments revealed that the elastic response
of individual double stranded DNA (dsDNA) molecules are excellently
described by the worm-like chain model introduced in polymer theory by
Kratky and Porod in 1949 \cite{RubCol03} (see later in Sec.~\ref{DNA})
. By pulling an individual molecule it was possible to experimentally
measure the force as a function of the molecular extension, also
called force-extension curve (FEC)
\cite{SmiFinBus92,BusMarSigSmi94}. These results verified the
prediction of the worm-like chain model for the FEC and provided the
first direct mechanical estimate of the persistence length of
individual DNA molecules (roughly speaking the persistence length is
the distance along the contour length of the molecule where the
molecule keeps a straight direction due to its bending rigidity) which
was in agreement with previous light scattering measurements. By
stretching the molecule above 60 pN ($1 \rm{pN}=10^{-12}N$) it was also
possible to observe a plateau in the FEC at 65 pN (see
Fig.~\ref{fig3}C). This is characteristic of a first-order transition
and interpreted as a structural change in the DNA molecule that gets
overstretched as the DNA double helix unwinds and bases get tilted in
the direction where the force is applied. The FEC of DNA (or any other
polymer) in single-molecule pulling experiments is similar to the
magnetization-field curve in magnets, the load-deformation curves in
plastic materials or the polarization-voltage curves in dielectrics
showing the physical flavor of such experiments.

A particular area of physics that can largely benefit from knowledge
emerging from SME is statistical physics. SME allow one to measure forces
in the range of a few piconewtons (a piconewton is approximately a
trillionth of the weight of an apple) and have spatial resolution in
the range of a nanometer. These are the ranges of forces and extensions typically
involved in many biomolecular reactions where high energy bonds (such
as ATP) are hydrolyzed and the energy released is subsequently used to
perform mechanical work. Work values typically encountered in such
reactions are of the order of a few $\rm{k_BT}$ units. At room
temperature $\rm{T}\simeq 300^\circ\rm{K}$ and therefore
$1 \rm{k_BT}\simeq 4~\rm{pN}\times{\rm nm}\simeq 0.6~\rm{kcal/mol}\simeq 2.5~\rm{kJ/mol}$
(In the biophysics or molecular biology community it is common to
refer to energies in $\rm{k_BT}$ or $\rm{kcal/mol}$ units whereas
$\rm{kJ/mol}$ is preferred among the chemists). Most biomolecular
reactions take place in an aqueous environment in the presence of
water molecules. As the energy of the biochemical reaction is not much
different than the average kinetic energy carried by one water
molecule such processes take place in a highly noisy
environment. Therefore, we can imagine a molecular enzyme acting on a
substrate carrying out a specific molecular reaction impinged by
hundreds of water molecules each nanosecond, each of these molecules
carrying enough kinetic energy to interfere in the process. Under such
conditions we expect strong Brownian fluctuations in the behavior of
the enzyme, which show up as rare and large deviations of its motion from
the average behavior \cite{FreKro05}. What if one water molecule
carrying 10 times the average kinetic energy collides with the enzyme
while a particular biochemical reaction takes place?  This happens
from time to time and such large deviations must influence the
behaviour of the motor. It is surprising to know that most of these
work producing molecular machines have a very large efficiency where a
large fraction of the energy consumed (typically around 20-90$\%$) is
used as mechanical work, part of the rest of energy gets lost in the
form of heat released into the aqueous environment.

Fluctuations are well known to be the cause of some mutations that occur
during the replication processes of DNA, when a new strand is
synthesized from the parental strand and the genetic information is
transmitted to a new generation of cells. During the replication process
many proteins interact with the DNA and participate in the process by
self-assembling into a large complex. The replicating machinery is
immersed in water and subjected to strong Brownian fluctuations. Under
such harsh conditions mutations occur frequently during the replicating
process. However, what mostly surprises from the replication process is
not the fact that mutations are common but that mutation rates are
regulated inside the cell. Mutations are necessary as evolution takes
advantage of them to produce better adapted individuals. However, mutation
levels during the replication process are kept under so tight control
that mismatches occur as rarely as one every $10^9$ replicated base
pairs. To avoid the damaging effects of noise fluctuations, cells are
endowed with a complex machinery of repair that is active during the crucial steps
of the replicating process and important for the maintenance of the
genome.  Large fluctuations are expected to occur whenever a large
number of fast moving molecules clash simultaneously with the
enzyme. Will these large deviations affect the performance of the
enzyme? In which way will they alter its function?  Even more
interesting, are these deviations an integral part of the function and
efficiency of the enzyme? Such questions are just a few among many
others that biophysicists and statistical physicists are ready to
confront.

The possibility to measure such tiny energies brings us close to what we
can distinguish as an emerging area of science, i.e. how to understand
and design molecular motors that operate in the nanoscale and
efficiently use chemical energy from naturally available (or
synthesized) sources to perform specifically designed functions. We
might call this thermodynamics of small systems as the main goal of this
discipline is to understand the performance of these small machines in a
noisy environment. Such term was already coined by Hill many years ago
when referring to thermodynamic equilibrium properties of ensembles of
small size systems where the equations of state depend on the ensemble
or collectivity, i.e. the conditions or parameters that are kept fixed
in the ensemble \cite{Hill94}. However, the most important aspect of
molecular machines is that they operate far from equilibrium by
hydrolyzing sizable amounts of energy and taking advantage of large and
rare deviations. The relation between the nonequilibrium properties of
small machines and their thermodynamic properties is shaping a new
discipline in statistical physics, the so called {\em nonequilibrium
thermodynamics of small systems}.  The main facts behind this new
discipline have been discussed in \cite{BusLipRit05} at a very
introductory level.

A related aspect of great interest to the statistical physicist is how
to use SME to test the basic foundations of statistical mechanics. This
line of thought has seen important progress in recent years and we
foresee that the exploration will continue and improve as more precise
and quantitative measurements are becoming available. Examples are the
study of the nonequilibrium work relations or fluctuation theorems under
various conditions. Biological systems have been particularly useful in
this regard. The experimental access to small energies is interesting in
non-linear systems, i.e. systems that do not respond in a linear way to
an applied external perturbation. Systems of this type abound in
molecular biology where conformational changes and macromolecular
interactions are of the all-or-none type. This is also referred as the
{\em lock and key} interaction mechanism among molecular biologists,
{\em substrate-enzyme} reactions in biochemistry or {\em activated behavior} in the
language of the physicist. In these type of interactions, a tiny
variation of the external conditions can cause a big change in the
outcome of the reaction. This fact is behind the high sensitivity of
protein structures to single amino acid mutations or the strong
dependence of some enzymatic reactions to a small amount of some
specific substances (e.g. activators or repressors). In proteins,
although not all single amino acid changes lead to new folded structures,
a few of them in some specific parts of the chain can have dramatic
effects in the structure with lethal consequences at the level of the
cellular functions regulated by that protein. In physical systems such
stability conditions are not so usually encountered. Upon the influence
of small perturbations physical systems usually respond in a smooth way
albeit, of course, counterexamples also abound. The main difference
between physical and biological systems is that structure and function
are intimately related in the latter. As a result of evolution over
millions of years a new kind of interactions and interrelationships have
emerged between different parts of living matter (evolutionary
constraints) which are not expected to be common in other physical or
chemical systems.

The idea of following individual biomolecules to understand processes
that occur in a crowded environment like the cell pertains to the kind
of abstractions typical of a physicist. How has been received this idea
by the community of biologists? The response is not yet uniform among
the rows of biologists who, depending on their area of specialization,
can feel more akin to the new techniques.  Molecular biologists and
biochemists are probably the most receptive because single-molecule
methods offer complementary tools to investigate problems of their
interest. For example, many processes that occur inside the cell such as
DNA transcription and replication, molecular transport, virus infection,
DNA condensation, ATP generation can be studied by using these new
techniques. As we discussed in Sec.~\ref{why}, this approach provides
new relevant information to the molecular biologist which is usually
unavailable with traditional techniques. A common criticism to
single-molecule methods is that molecular processes cannot be studied in
vivo by following a molecule in its natural environment (e.g. a DNA
replicating inside the nucleus). This is indeed a limitation at present
but it must be said that the same occurs in traditional biochemical
assays. The specific conditions found in the cell cannot be reproduced in the
test tube, which contains only a tiny fraction of the total number of
cell constituents.  This limitation, however, is not seen as a drawback
in the mind of a physicist who is educated to explore simplified
versions of complex and difficult problems. The true fact is that more
and more molecular biologists are becoming steadily interested in
adopting such techniques in their labs. This gives rise to an
unprecedented excitement between physicists and biologists who are
joining efforts and expertise to accomplish common scientific goals.

\section{Experimental techniques}
\label{exptech}

In this section I succinctly describe the experimental techniques mostly
used in SME. We have to distinguish between techniques required to
manipulate individual molecules and techniques that allow to detect and
follow in real time (but not manipulate) individual molecules. In the
first class we have atomic force microscopy (AFM), laser optical tweezers (LOT), magnetic tweezers
(MT) and biomembrane force probe (BFP) to cite the most representative
ones. Other techniques (such as glass microfibers) are not of widespread
use in SME and we are not going to discuss them here. In the second
class, and in addition to AFM (which is also an imaging technique),
there are predominantly optical techniques such as single-molecule
fluorescence (SMF), Raman spectroscopy, two-photon spectroscopy and
semiconductor quantum dot emission to cite the most common. Combination
of manipulation and fluorescence techniques (e.g. laser tweezers with
fluorescence) has already begun and will allow to explore new phenomena
with enhanced precision.

\begin{figure}
  \centering
  \includegraphics[scale=.55,angle=-90]{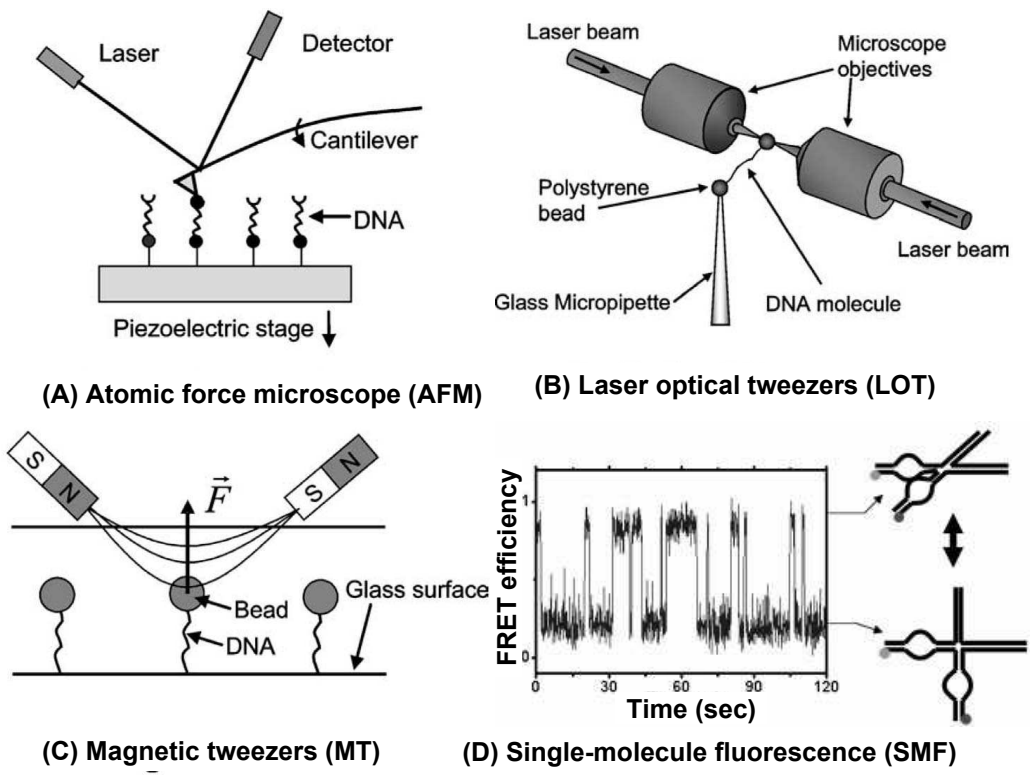}
  \caption{Experimental techniques: (A) Atomic force microscope
  (AFM); (B) Laser optical tweezers (LOT); (C) Magnetic tweezers; (D)
  Single-molecule fluorescence (SMF) using FRET.}
  \label{fig2}
\end{figure}

\subsection{Single-molecule manipulation techniques}

In this section the different techniques to manipulate individual
molecules and measure microscopic forces are outlined. AFM are
particularly useful because these can be used to sweep surfaces and
take images of individual molecules. At the same time, AFM can be used
to apply mechanical force on individual molecules. LOT cannot be used
to take images of individual molecules but have the advantage that
manipulation of individual molecules can be more easily controlled,
for example, in the study of molecular motors. The case of LOT has
been chosen to be discussed in detail to emphasize the difficulties
and the methodology common to all techniques. All techniques discussed
in this section have complementary forces and loading rate regimes,
each technique enables some experiments that are not possible with the
other. Most of the methods described to capture video images and other
details of the experimental LOT setup (such as the implementation of
force-clamp methods) are also applied to the other techniques.

\subsubsection{Atomic-force microscopy (AFM).}
\label{afm}

Probably the best well known among these techniques is atomic force
microscopy (AFM). The AFM is a descendant of the scanning tunneling
microscope invented by G. Binnig and H. Rohrer applied to obtain
Angstrom resolution images of metallic surfaces using the quantum tunnel
effect \cite{BinRoh87}. The AFM \cite{BinQuaBer86} is based on the
principle that a very soft cantilever with a tip that is moved to the
vicinity of a surface (metallic or insulating) can sense the roughness
of the surface and deflect by an amount which is proportional to the
proximity of the tip to the surface (Fig.~\ref{fig2}A). The most
important application of the AFM is imaging where it can work
in various modes: the contact mode, tapping mode and the jumping
mode \cite{ColGomBar04}. For example, in the tapping mode the tip is made to oscillate
close to the sample surface. The amplitude of the oscillation is
recorded and controlled by a feedback loop mechanism that keeps such
amplitude constant. When passing over a bump the amplitude decreases so
the distance between tip and surface is increased to keep the amplitude
of oscillation constant. When passing over a depression the tip is moved
to the surface. This mode has the advantage that the transverse motion
of the tip along the surface is not influenced by shearing and
frictional forces thereby avoiding damage to the sample and noisy
interference effects.  A map of the distance of the tip to the sample
provides an accurate topographic image of the surface. Other modes are preferable
depending on the particular system, for example the contact mode is
useful to take images of biological samples in fluids \cite{HorMil03}. The use of the
AFM for biomolecular imaging has been reviewed in several excellent papers
\cite{BusRivKel97,CzaSha98,Smith00,HafCheWooLie00,EngMul00,MulAnd02,Kel05}.

The AFM is also used to manipulate and exert mechanical force on
individual molecules  (Fig.~\ref{fig2}A). As usual in single molecule tecniques, elaborated
chemistry is often required to treat the surface and the tip. The
surface has to be coated with the molecules to be manipulated. The AFM
tip also has to be coated with molecules that can bind (either
specifically or non-specifically) to the molecules on the substrate. By
moving the tip to the substrate a contact between the tip and one of the
molecules adsorbed on the substrate is made. Retraction of the tip at
constant speed allows to measure the deflection of the tip in real time
providing a measure of the force acting on the molecule as a function of
its extension, the so-called force-extension curve (FEC).  The AFM
covers forces in the [20 pN-10 nN] range depending on the stiffness of the
cantilever. Typical values of the stiffness are in the range
$10-1000$ pN/nm. Although AFM is a very versatile and powerful tool it
has a few drawbacks for manipulating single molecules. The most
important one is probably the presence of undesired interactions between
tip and substrate (Van der Waals, electrostatic and adhesion forces) and
the non-specificity of the attachments that often occur between tip and
substrate. When moving the tip to the substrate it is easy to attach
many molecules at a time. Moreover, it is difficult to control the
specific location of the attachment between the tip and the molecule.
Single molecule markers (e.g. polyproteins) and functionalization
strategies have been specifically developed to overcome these limitations.

Spatial and force resolution in the AFM are limited by thermal
fluctuations. When the cantilever stage is held at a constant position
the force acting on the tip and the extension between tip and substrate
fluctuate. The respective fluctuations are given by $\langle \delta
x^2\rangle=k_BT/k$ and $\langle \delta F^2\rangle=k_BT k$ where $k_B$ is
the Boltzmann constant, $T$ is the absolute temperature of the
environment and $k$ is the stiffness of the cantilever.  At room
temperature $k_BT\simeq 4$ pN$\times$nm and therefore $\sqrt{\langle
\delta F^2\rangle}\simeq$ 20 pN,$\sqrt{\langle \delta x^2\rangle}\simeq
$ 2 \AA~ if we take $k\simeq 100$ pN/nm. This shows that the
signal-to-noise ratio for the force is small for force values of just a
few tens of pN. This is the range of forces characteristic of weak
interactions therefore showing the limitations of AFMs to study the
mechanochemistry of weak interactions in the lower pN regime.  In
contrast, AFMs are ideal to investigate strong to covalent
interactions. They have been mostly used to probe relatively strong
intermolecular and intramolecular interactions, e.g. pulling
experiments in biopolymers such as polysaccharides, proteins and nucleic acids.

\subsubsection{Laser optical tweezers (LOT).}
\label{lot}
The principle of LOT is based on the optical gradient force generated by
a focused beam of light acting on an object with an index of refraction
higher than that of the surrounding medium. Discovered by A. Ashkin in
1970 \cite{Ashkin70a,Ashkin70b} the principle was developed more
recently to trap dielectric particles by A. Ashkin and collaborators at
the Bell Labs \cite{AshDziBjoChu86}. The application of gradient force
by light radiation pressure has been used to trap neutral atoms
\cite{Phillips98}, eukaryotic cells \cite{AshDziYam87} and virus and
bacteria \cite{AshDzi87}.  A good review on the origins of optical
trapping can be found in \cite{Ashkin97}. In the basic experimental LOT
setup a near-infrared laser is collimated by a high numerical aperture
water immersion lens. A micron-sized polystyrene or silica bead is then
trapped in the focus of the laser by exerting forces in the range
0.1-100 pN depending on the size of the bead and the power of the
laser. Typical bead sizes are on the order of 1-3 microns and laser
powers of a few hundreds of milliwatts to avoid the heating of the bead
and undesired heat convection effects close to the bead that could
either damage the sample or affect the measurements.  To a very good
approximation the trapping potential is harmonic, therefore forces
acting on the bead are directly proportional to the distance between the
bead and the center of the trap, $F=kx$ where $k$ is the stiffness
constant of the trap. To determine the stiffness of the trap, noise
measurements or Stokes force calibration are often used. Typical values
of the stiffness of the trap are $10^2-10^4$ times smaller than AFM
tips, therefore force resolution is at least 10 times better on the
order of 0.1 pN.  Major improvement in this basic setup is obtained by
using dual counter propagating laser beams passing through two identical
objectives \cite{SmiCuiBus03} (Fig.~\ref{fig2}B). There are several
advantages in this more complex setup. First, the axial scattering force
is reduced. Second, the trapping forces that can be reached (up to
200 pN) are higher than in the one-beam setup. Finally, continued force
calibration is not required because the force is directly measured from
the total amount of light deflected by the bead which is collected by
position sensitive detectors (PSD) located at the two opposite sides of
the laser beams. The force is given by $F=S/(Lc)$ where $S$ is the
radiation pressure flux, $c$ is the speed of light, $L$ is the distance
that separates the bead (located between the two objectives) and one of
the objectives. This formula is valid for any size, shape and refractive
index of the bead. In this way there is no need for the calibration of
the trap every time a new bead is captured.

\begin{figure}
  \centering \includegraphics[scale=.5,angle=-90]{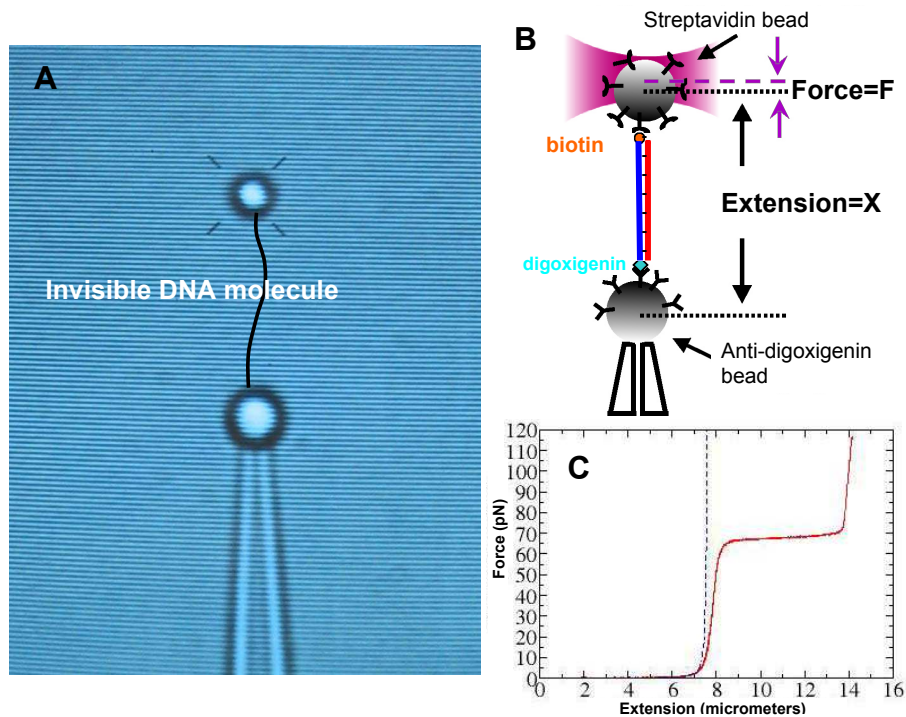}
  \caption{Pulling DNA using LOT. (A) Image taken by a CCD camera of
  the fluids chamber where manipulation of the DNA molecule takes
  place. The DNA is tethered between two beads. The upper bead is
  trapped in the optical well whereas the lower bead is immobilized on
  the tip of a micropipette and held fixed by air suction. The
  presence of a DNA tether between the beads (invisible in the real
  image but illustrated here as a thick black line) is detected by the
  presence of a vertical force pointing downward acting on the trapped
  bead. (B) Experimental setup in DNA pulling experiments where the 3'
  and 5' ends of one strand are attached to the beads through
  biotin-streptavidin and digoxigenin-antidig specific bonds. The
  force is then measured from the deviation of the upper bead respect
  to the center of the trap. By measuring the force (F) as a function
  of the extension (X) it is possible to record the force-extension
  curve (FEC). (C) FEC in half $\lambda$-DNA (24000 base-pairs
  long). The blue-dashed line is the worm-like chain prediction from
  polymer theory. It describes very well the elastic behavior of the
  DNA molecule up to forces $\sim$ 5 pN. Above 5 pN enthalpic
  corrections are important. Note the overstretching transition that
  occurs at 65 pN. See the more detailed discussion in Sec.~\ref{DNA}.}
  \label{fig3}
\end{figure}

Manipulation of individual molecules is carried out inside a fluids
chamber made out of two glass surfaces separated by a parafilm of 200
microns width. A fluidics system is designed in such a way that water
and chemicals can be flowed and replaced at any time inside the
chamber. A glass micropipette is inserted and fixed inside the chamber
and used to hold another bead by air suction. The chamber is then
mounted on a moving stage whose position is controlled by a piezo and
positioned in front of the objective lens so that the laser can be
focused inside the chamber. To manipulate individual molecules beads are
coated with a chemical substance (e.g. avidin or streptavidin) that can
bind specifically to its complementary molecule (biotin). The molecules
of interest (e.g. DNA) are then biotinylated (i.e. labeled with biotin
molecules) at their ends so they can bind to the avidin (streptavidin)
coated beads through a strong non-covalent bond. To avoid double
attachments between the two ends of a single molecule and the same bead
it is customary to differently label the molecule at its two ends. One
end is then labeled with biotin, the other with digoxigenin. Digoxigenin
recognizes specifically its anti-dig antibody partner through the "lock
and key" interaction mechanism typical of antigen-antibody
interactions. This is a weak bond so the biomolecule is often labeled
with many dig molecules at one of its ends in order to increase the
strength of the attachment. After incubation of the molecules with the
beads (for instance the streptavidin beads) other beads coated with
anti-dig are flowed inside the chamber. One bead is captured with the
laser trap and moved to the tip of the micropipette where it is held
fixed by air suction. Incubated beads are then flowed inside the chamber
and one bead is captured in the trap. By moving the chamber relative to
the trap the two beads approach each other until a connection between
the digoxigenin end of the molecule and the bead in the micropipette is
established. The tether is then pulled by moving the chamber at a given
speed and the FEC measured, see Fig.~\ref{fig3}.

The extension of the molecule can be monitored by using a CCD video
camera that uses a framegrabber to take pictures of the two beads and
operates at a few tens of a Hertz. Because spatial resolution is
strongly limited by the pixel resolution (about 10 nm) other methods have
to be implemented to resolve the position of the bead with higher
accuracy.  A standard procedure is to use a light lever or reference
beam where a low power light beam passing through a small lens in the
frame of the chamber is collected by using additional position sensitive
detectors. By recording the position of the chamber it is possible to
determine the extension of the tether down to a few nanometers of
precision. Spatial resolution is however hampered by strong drift
effects in the optical components and the manipulation
chamber. Depending on the experimental setup the spatial resolution can
reach the nanometer only in carefully isolated environments (absence of
air currents, mechanical and acoustic vibrations and temperature
oscillations).  Force-clamp (also called force-feedback) methods that
use acoustic-optic deflectors and incorporate a piezoelectric stage with
capacitive position sensing are providing better and more versatile
instruments \cite{VisBlo98,LanAsbShaBlo02}.  LOT have been widely used
to investigate nucleic acids and molecular motors.

\subsubsection{Magnetic tweezers (MT).}
\label{mt}

Magnetic tweezers (MT) design is based on the principle that a
magnetized bead experiences a force when immersed in a magnetic field
gradient $F=-\mu \nabla B$. The basic setup is shown in
Fig.~\ref{fig2}C. A bead is trapped in the magnetic field gradient
generated by two strong magnets. Molecules are attached to the surface
of the magnetic bead and to a glass surface. A microscopic objective
with a CCD camera is used to determine the position of the bead.
Molecules are pulled by moving the translation stage that supports the
magnets.  MT has several advantages compared to AFM and LOT
\cite{GosCro02}. First, sensitivity to very low forces can be easily
achieved due to the low value of the stiffness of the magnetic trap. The
typical range of forces is $10^{-2}$ pN-10 pN where the maximum value of
the force depends on the size of the magnetic bead. Second, in the
passive mode (when the magnets stage is kept fixed) the force acting on
the bead can be kept constant because the spatial region occupied by the
bead is small enough for the magnetic field gradient to be considered
uniform. Therefore, although the bead position fluctuates the force is
always constant. A constant force can be also achieved in the AFM and
LOT setups by implementing force-feedback control mechanisms (see
Sec.~\ref{lot}). The main drawback of feedback loops is their working
frequency, typically limited to a few KHz, which do not allow to detect
dynamical processes faster than milliseconds. Third, magnetic traps
allow to twist molecules by rotating the magnets. Modifications of the
basic setup by using a third bead to create a single chemical bond
swivel allow also to measure torques \cite{BryStoGorSmiCozBus03}. MT are
calibrated in flow fields using the Stokes' law or measuring Brownian
motion in the direction transverse to the application of the force,
$k=K_BT/\langle \delta x^2\rangle$ where $x$ denotes the transverse
coordinate. Typical values of the magnetic trap stiffness are
$10^{-4}$ pN/nm thereby one million times smaller than in AFMs and a
thousand times smaller than in LOT. Force-extension curves (FECs) can be
recorded in real time by moving the stage and measuring the transverse
fluctuations $\langle \delta x^2\rangle$. Force is measured by using the
expression $F=k_BT l/\langle \delta x^2\rangle$ where $l$ is the
extension of the molecule. The value of $l$ is determined using depth
imaging techniques that provide excellent force-position measurements. The
smallness of the stiffness in MTs induces large fluctuations in the
extension of the molecule on the order of 20 nm. MTs have been
extensively used to investigate elastic and torsional properties of DNA
molecules.

\subsubsection{Biomembrane force probe (BFP).}
\label{bfp}

\begin{figure}
  \centering
  \includegraphics[scale=.8,angle=0]{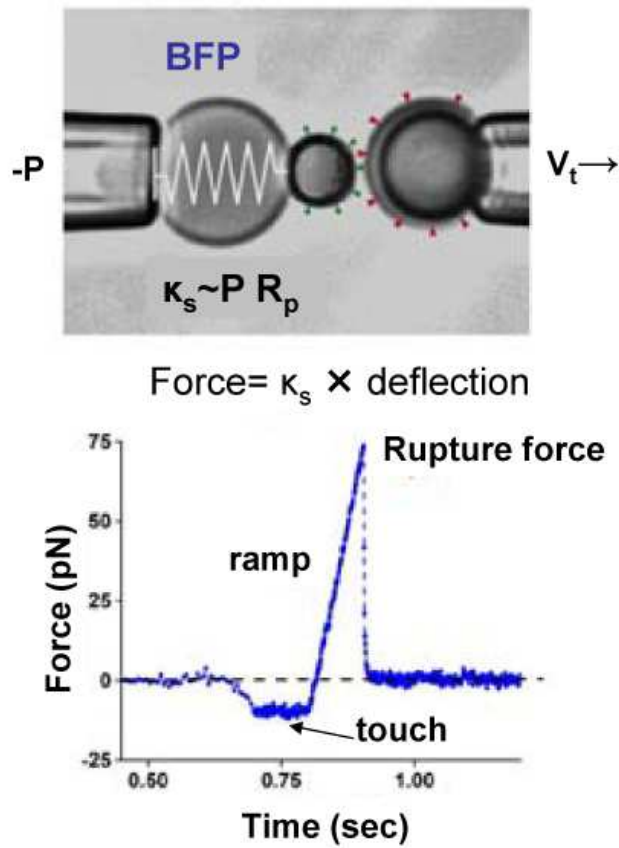}
  \caption{BFP technique. (Upper figure) A red blood cell acts as a force transducer
  (symbolized as a spring in the figure) by transforming the pressure
  suction applied on the
  pipette (-P) into the elastic stiffness of the cell membrane $\kappa_s\sim P
  R_p$ where $R_p$ is the pipette radius. A bead covered with ligands is
  then attached to the cell (left bead). Another bead covered with receptors is then
  immobilized on the tip of another pipette (right bead). A ligand-receptor bond can
  be formed by touching the beads. By retracting the right pipette at
  speed $v_t$ the formed bond dissociates at a given force value $F$, measured
  from the formula $F=\kappa_s x$ where $x$ is the deflection of the left
  bead. (Lower figure) Typical force-time curve when probing ligand-receptor
  interactions. Figure taken from \cite{EvaWil00}.}
  \label{fig4}
\end{figure}

Finally, we mention the biomembrane force probe technique developed
by Evans and collaborators \cite{MerNasLeuRitEva99}. The basic
experimental setup is shown in Fig.~\ref{fig4}. In this setup a
biotinylated red blood cell is pressurized by micropipette suction
into a spherical shape. The tip is made of a streptavidin coated bead
(left bead in Fig.~\ref{fig4}) functionalized with some molecules
(e.g. ligands). The other bead is functionalized with complementary
molecules (e.g. receptors) and kept fixed by air suction on the tip of the other
micropipette (right). The blood cell acts like a spring so the force
can be measured by calibrating the stiffness of the cell. This is
directly related to the membrane tension and can be controlled by fine
tuning of the pressurization of the micropipette (left). Typical
stiffness values are in the range 0.1-1 pN/nm.  By moving the
micropipette (right) using a piezo translator stage the two beads
approach each other until they touch. The micropipette is further
retracted and ligand-receptor interactions detected as rupture
events. A CCD video camera records the extension between the two beads
with a resolution on the order of 10 nm.  The BFP has been mainly used
to study ligand-receptor interactions.

\subsection{Single-molecule fluorescence techniques}
\label{smf}

Single-molecule fluorescence (SMF) is based on the detection of light
emitted by fluorophores that have been attached to the molecule under
quest. Fluorophores are excited from their ground state by absorbing
light from an external light source. After internal conversion and
vibrational relaxation they emit fluorescent light in $10^{-9}-10^{-7}$
seconds.  Detection of single molecules is possible by exciting a very
small volume with light and observing the emitted signal. Typical
volumes are on the order of the femtoliter ($10^{-15}$ l) corresponding
to a water drop of diameter on the order of a micron.  In diluted
solutions (on the order of nanomolar concentration) the typical number
of photons intercepted by a molecule are estimated to be on the order of
a million photons per second giving a number of photons emitted in such
volume of a thousand photons per second.  The emitted light can be detected
using sensitive photodetectors such as avalanche photodiodes and
photomultipliers. The main advantage of SMF is its high-time
resolution. This covers from the range of microseconds for individual
events up to picoseconds when identical experiments are repeated many
times and lots of statistics are collected. SMF is a non-invasive
technique that can be used to study biological samples in vivo.  In
general the spatial resolution in an optical system is limited by the
Rayleigh criterion (typically around 200 nm). However the limit imposed
by the diffraction of light can be overcome by using other methods such as
FIONA, a method based on centroid localization where nanometer precision
can be achieved when a sufficient number of photons is collected
\cite{GelSchShe88,GorHaSel04,QuWuMetSch04}. Other methods include SMF polarization
measurements using total internal reflection fluorescence (TIRF) that
allow to determine the orientation of individual molecules with tens of
millisecond resolution \cite{ForQuiShaCorGol03}.  With these methods it
has been possible to determine that myosin V, a two-headed molecular
motor that translocates along actin filaments, walks in a hand-over way by
alternating the role of the lead and trail heads during its motion
\cite{ForQuiShaCorGol03,YilForMcKHaGolSel03} (see Sec.~\ref{motors}).

A powerful SMF technique to detect conformational changes within 10 nm
is fluorescence resonance energy transfer (FRET) discovered by
F\"orster (sometimes called F\"orster resonance energy transfer) and
discussed already in Sec.~\ref{why}. FRET is based on a quantum
interaction mechanism between two fluorescent dyes that can transfer
energy when they are kept close enough (typically within a distance of
10 nm) and one of them (the so called donor) is excited by light. When
the donor is excited by an external source of light the other dye (the
so called acceptor) emits part of the light at another wavelength
through a non-radiative resonance energy transfer mechanism between
donor and acceptor. The efficiency of energy transfer depends on the
distance between donor and acceptor according to the formula
$E=E_0/(1+(R/R_0)^6)$ where $R_0$ is F\"orster radius or the value
of the distance above which the efficiency goes below $50\%$. The
value of $R_0$ is a function of the fluorescence and absorption
spectrum of donor and acceptor and the orientation between the
electric dipoles of both molecules. F\"orster efficiency formula gives
a spectroscopic ruler to determine distances between the two dyes
thereby allowing to identify conformational transitions in the
biomolecule (Fig.~\ref{fig2}D).

Difficulties associated to SMF are the expertise required to chemically attach
fluorophores in biomolecules where it often consists of a "try and
repeat" procedure. FRET has the added complication that two dyes have to
be chemically attached to the same molecule at specific locations. 
Often it is not possible to know the dipolar orientation of
the dyes which makes difficult to determine their distance using the
spectroscopic ruler. Combination of FRET with other techniques (e.g. electron
microscopy or X-ray diffraction) helps to identify and characterize
conformational changes. A widespread problem in SMF is photobleaching of
fluorophores. Photobleaching is a process by which excited fluorophores
undergo a chemical transformation (e.g. after reacting with oxygen) and
stop fluorescing. Methods are currently employed to reduce such effect
which still is a main nuisance of SMF techniques.

SMF has been used to study molecular transport, protein folding and
conformational transitions in enzymatic reactions. There is much current
effort to combine SMF with force measurements. This would allow to
identify conformational changes with force jumps thereby giving precious
information about biomolecular function.

\section{Systems}

Single-molecule techniques have been applied to a great variety of
systems. From polymers to living cells many system properties have been
characterized and studied. In what follows I provide a general
overview of a personal selection of these problems. My choice is unavoidably biased
by my specific knowledge of some of these questions.

\subsection{Nucleic acids}

\subsubsection{DNA.}
\label{DNA}

Historically DNA is the most important player in molecular biology
\cite{nature74,nature03}. Since the discovery of the double helix in
1953 \cite{WatCri53} the structure of the DNA molecule has been
studied under different conditions using crystallographic and bulk
methods \cite{Kamenetskii97,Tinoco96}. DNA can be found in various
structural forms and it is now recognized that the phase diagram of
the molecule in the presence of force and torque is rich and
complex. Pulling experiments in DNA \cite{Marko97} use glass
microfibers \cite{CluLebHelLavVioChaCar95}, LOT \cite{SmiCuiBus95},
MTs \cite{StrAllBenBenCro96} or AFM \cite{RieClaGau99}.  Pioneering
work in the study of the elastic properties of DNA was carried out by
Finzi, Smith and Bustamante in 1992 when they visualized the motion of
fluorescent DNA molecules attached to micron-sized beads acted by
magnetic and hydrodynamic forces \cite{SmiFinBus92}. B-form DNA
molecules 48 Kbps long (one base pair -bp- is about 3 \AA~ long) of the
$\lambda$-bacteriophage virus were stretched up to forces as high as 100 pN
using AFM \cite{RieClaGau99} and LOT \cite{SmiCuiBus95}.  DNA shows a
elastic response at low forces (below 5 pN) dominated by entropic
effects whereas at high forces (above 5 pN) enthalpic contributions
start to be important. Pulling experiments in DNA confirmed that B-DNA is an elastic
molecule whose force-extension behavior can be well described using
the worm-like chain model
\cite{BusMarSigSmi94,Vologodskii94,MarSig95,WanYinLanGelBlo97,BouWanAllStrBloCro99}
introduced in polymer theory by Kratky and Porod
\cite{RubCol03}
\footnote{The solution of the model under the action of external force
is equivalent to the solution of the classical Heisenberg
ferromagnetic chain in a magnetic field \cite{GurSca75}.}. Above 5 pN
the FEC is well described by the phenomeological extensible worm-like
chain where the contour length $L$ changes as a function of the
applied force $F$ by $\Delta L=LF/Y$ where $Y$ is the Young elastic
modulus ($Y\simeq 1~{\rm nN}$). In torsionally unconstrained DNA one end is
immobilized in one bead, the other end of either one of the two
strands (3' or 5') is immobilized to another bead or surface
(depending on the experimental setup). For torsionally unconstrained
DNA a transition is identified at an applied force of 65 pN where the
B-DNA overstretchs into a new form (the so called S-form DNA or S-DNA)
where the new extension of the molecule is approximately 1.7 times its
original contour length. In the S form the double helix
(characteristic of the B form) unwinds and all base pairs tilt along
the force direction. The overstretching transition has a force plateau
characteristic of first-order phase transitions (see
Fig.~\ref{fig3}C). Curiously enough this extended form of DNA was
anticipated 50 years ago from the measurement of the optical
properties of fibers under tension \cite{WilGosSee51}. A first order
transition is also found when DNA melts when heated up above
65$^\circ$C. Although it has been suggested that S form is
force-induced melted DNA \cite{RouBlo01a,RouBlo01b,WilRouBlo02} the
current evidence suggests that S-DNA keeps the Watson-Crick base pairs
intact in the absence of nicks along the DNA phosphodiester
backbone. Noticeable salt dependence effects have been reported for
the elastic properties and overstretching transition in DNA due to its
large electrostatic charge \cite{WenWilRouBlo02}. Various statistical
models have been introduced in the literature that investigate several
aspects of DNA such as thermal denaturation
\cite{PolShe66a,PolShe66b,WarBen85,PeyBis89,CulHwa97,KafMukPel00,CauColGra00,SalGuiMorWidAma05,CuePeyGra05}
in the presence of force and torque \cite{CocMon99,CocMon00}, bubble
formation \cite{HwaMarSneTan03,HanMet03} and the overstretching
transition
\cite{KonBol96,AhsRudBru98,KosGorZhuOls99,StoNel03,MerEjeEve03,CocYanLegChaMar04}. The
response of DNA to mechanical force is similar to that observed in
other biopolymers such as peptides and polysacharides. These show an
elastic response at low forces which is dominated by entropic effects
whereas at high forces enthalpic contributions and structural
transitions are often observed
\cite{RieOesHeyGau97,MarOberPanFer98,MarLiObeFer02,KawByrKhaLeiRadSmi04,HugRieSeiGauNet05}.

\begin{figure}
  \centering \includegraphics[scale=.6,angle=-90]{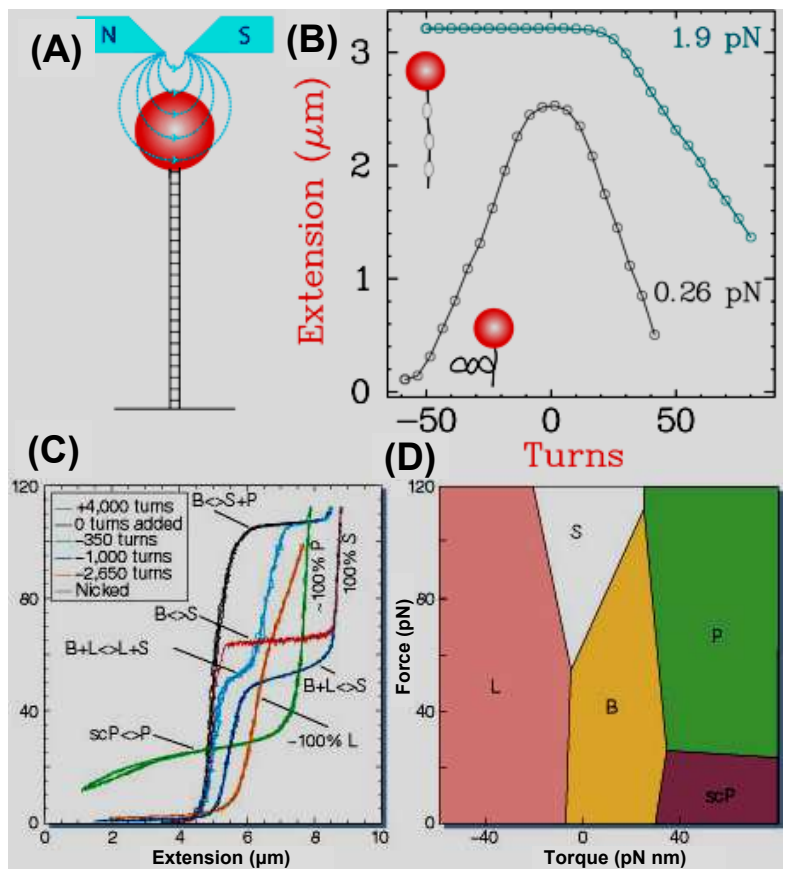}
  \caption{Force-torque DNA measurements. (A) Experimental setup with
  MT. DNA can be twisted by exerting a torque through the rotation of
  two magnets. (B) Extension-twist curves showed a marked different
  behavior depending on the value of the constant applied force. If
  undertwisted the molecule forms bubbles at high forces, whereas it
  forms plectonemes at low forces. (C) Various FECs obtained by pulling
  on twisted DNA using LOT. The different plateaus indicate different
  structural transitions. The possible DNA forms are: normal B-form DNA,
  overstretched (S-DNA), the highly overwound DNA Pauling (P) form,
  supercoiled and shortened Pauling (scP) form, and the underwound and
  denatured (L) form. (D) Phase diagram indicating all possible phases.
  Pictures (A,B) were taken from
  \cite{DekRybDugCriCozBenCro02}. Pictures (C,D) were taken from
  \cite{BrySmiBus03}.}  \label{fig5}
\end{figure}

Current experiments can now exert force and torque at the same time on torsionally
constrained DNA. In torsionally constrained DNA both strands are
immobilized at one end of the molecule. MT allow to rotate a
magnetized bead attached to the end of a DNA molecule that is attached
to a glass coverslip through its other end
\cite{StrAllBenBenCro96}. DNA is a coiled molecule of 2 nm diameter
covering one helical turn every 10.5 bps equal to 3.4 nm, also called
helical pitch. By exerting torque it is possible to change the helical
pitch of the molecule leading to a supercoiled molecule. The
topological properties of a closed DNA molecule are determined by the
so-called linking number Lk which is equal to the number of crossings
between the two strands. In a closed DNA molecule (such as circular
DNA from bacteria) Lk is a topological invariant equal to the sum of
twist (Tw) and writhe (Wr), Lk=Tw+Wr. The twist is the number of
helical turns whereas the writhe is the number of loops occurring
along the DNA molecule. The amount of supercoiling, usually termed
as superhelical density $\sigma$, is measured as
$\sigma=\frac{Lk-Lk_0} {Lk_0}$ where $Lk_0$ is the linking number of
torsionally relaxed DNA. Supercoiling is a biological important
property of DNA.  Supercoiling plays an active part in the regulation
of the genome in both eukaryotes and bacteria (inside cells DNA is
negatively supercoiled, $\sigma\sim -0.06$) and is controlled by a
family of enzymes called topoisomerases (see below in
Sec.~\ref{motors}). These are involved in packaging, transcription,
replication, repair and recombination of genomic DNA. Under the
application of constant force the extension of the DNA molecule
changes as the bead is rotated and the twist increases or decreases
(Fig.~\ref{fig5} (A-B)). Under twist various structural transitions
are observed \cite{LegRomSarRobBouChaMar99,SarLegChaMar01}.  At very
high forces a new form of DNA (called P-form) is found when the DNA is
overtwisted and the extension of the molecule decreases
\cite{AllBenLavCro98}. In this new form bases are extruded from the
inside of the backbone in a form that is reminiscent of the triple
strand model for the structure of DNA proposed by Pauling in the 50's
(henceforth the name P for the phase) \cite{PauCor53}. At low forces
the DNA forms plectonemic supercoils (the coils often observed in
telephone cords) if overtwisted and denatured bubbles if
undertwisted. All these behaviors have been extensively studied and
modeled
\cite{LebLav96,FaiRudOst97,Marko98,BouMez98,ZhoZhaYan00,Neukirch04}. They
have resulted into a complex force-torque phase diagram of DNA
\cite{BryStoGorSmiCozBus03,BrySmiBus03} (see Fig.~\ref{fig5} (C-D)).
Conformational fluctuations have been shown also to be important in
regulatory mechanisms such as protein-DNA interactions
\cite{LegRobBouChaMar98,PanKarWil03}.

In another class of experiments the 3' and 5' ends on one end of
the DNA molecule are immobilized into a bead and a surface. By pulling
apart the bead from the surface the Watson-Crick base pairs connecting
the two strands along the phosphate backbone break and the DNA molecule
opens like a zipper. Unzipping is the process where hydrogen bonds
between complementary bases fall apart and the bases that are buried
inside the helix become exposed to the solvent. It naturally occurs in
various biomolecular processes. For example, the initiation process
during DNA replication is led by the exposure of specific DNA segments
of the genome to the replicating machinery (a set of various multimeric
protein complexes). The replication of DNA is then determined by the
advance of the replication fork which proceeds by untwisting and
unzipping the DNA structure. Typical unzipping forces are on the order
of 15 pN and are sequence dependent
\cite{EssBocHes97,BocEssHes98,RieClaGau99,ClaRieTolGau00,DanColBouLubNelPre03}. Below
such force (e.g. around 10 pN) the molecule does not unzip. Above that
force (e.g. at 17 pN) unzipping proceeds fast with a signal that is a
fingerprint of the DNA sequence.  During the unzipping process, the
recorded force/extension signal does strongly depend on the particular
DNA sequence. By unzipping the molecule a repeated number of times a
characteristic force pattern emerges except for thermal
fluctuations. This fact makes force unzipping a promising technique for
DNA sequencing \cite{BhaMar02,CocMonMar01,CocMonMar02,CocMarMon06}.
Similar experiments are being also conducted with RNA and will be
discussed below in Sec.~\ref{RNA}. Force unzipping has been investigated
using various statistical models
\cite{LubNel02,Chen02,KafMukPel01,AllGevHuWu04} which predict re-entrance
of the transition due to excluded volume effects between the two strands
\cite{MukSha02,OrlBhaMarMarSen01,MarBhaMarOrlSen02}.  At present,
various technical precision limitations and noise fluctuations due to
the softness of the single stranded DNA (ssDNA) handles limit the reliability of the
sequencing procedure. Although at present it is possible to resolve the
unzipping of a DNA patch containing a few tens of base pairs, the
detection of the opening of just one base pair is an experimental
challenge. Improved detection resolution using two traps and combination
of LOT with SMF \cite{Ha01} are expected to provide far better accurate
results. Other sequencing strategies are SMF measurements of
DNA polymerase activity during the synthesis of the complementary strand
\cite{BraHebKarQua03}.

\subsubsection{RNA.}
\label{RNA}

RNA is a very important player in molecular biology. It participates in
most processes where the genomic information that is kept inside the
nucleus must be exported to the cytoplasm of the cell where it is
translated into proteins that are synthesized in the ribosome. The
ribosome, one of the largest biological machines in the cell, contains
RNA as part of its structure as well as ribosomal proteins. RNA is
considered a relic of the past \cite{Deduve95} where the precursors of
the first living cells (more than 2 billion years ago) used the
chemistry of RNA long before proteins took over most of the essential
functions of ancestral prokaryotic cells. The RNA world refers to a
hypothetical scenario where the majority of important living functions
were carried out by ancient RNA molecules \cite{Gestland93}. There are
many functions where RNA is essential (e.g. in enzymatic and regulatory
processes). Every few years new biological roles of the RNA are
discovered. RNA and DNA are chemically very similar molecules.  The main
difference of RNA with respect to DNA is the presence of the highly
reactive OH group in the sugar (ribose) and the replacement of thymine
by uracil (adenine pairs with uracil) in RNA. Structurally they are also
very different: RNA is found in nature in single stranded form whereas
DNA is found in double stranded form.  

The elastic properties of synthesized double stranded RNA and DNA are
very similar \cite{AbeMorHeiDekDek05}, but the single strand nature of
RNA makes the difference. RNA is a more complex molecule than
DNA. While complementariness is strict in DNA, RNA allows for
additional non Watson-Crick base pairs (such as GU or GA) between the
bases. Consequently it allows for more base pair interactions and a
larger number of possible structures. Single stranded RNA molecules
form complex secondary structures of stems, junctions, loops, bulges
and other motifs. The main thermodynamic stability of the molecule is
derived from the complementarity of the bases and base stacking
interactions.  However, secondary RNA structures fold into more
complex three-dimensional structures stabilized by specific
interactions between different hydroxyl groups of the bases that bind
divalent metal cations (e.g. magnesium). Tertiary interactions produce
other sort of structural elements such as pseudoknots or kissing
loops. In RNA the contribution to the free energy of the native state
due to the tertiary structure is a perturbation to the main
contribution due to the secondary structure. In addition, the
chemistry of RNA is much simpler than that of proteins (there are 4
different nucleotides in RNA as compared to the 20 amino acids in
proteins). This fact makes RNA easier to study at both theoretical and
experimental level. RNA research is very attractive to the
biophysicist not only because RNA is less complex than proteins, but
also because RNA shows all important properties exhibited by
proteins. RNA molecules can be unfolded under the action of mechanical
force \cite{TinBus02,Tin04,OnoTin04}.  Thermal and force denaturation
experiments reveal that RNA folds into a native three dimensional
structure in the same way proteins do. Understanding how RNA folds
will help to better understand the corresponding process in proteins
\cite{Herschlag95,TinBus99,SchBarSem04}.  The problem of RNA folding
has also motivated theoretical insight from the theory of disordered
systems in statistical
physics~\cite{Higgs96,PagParRic00,OrlZee02,BunHwa02,Muller02,MulKrzMez02}.

\begin{figure}
\centering   \includegraphics[scale=.5,angle=-90]{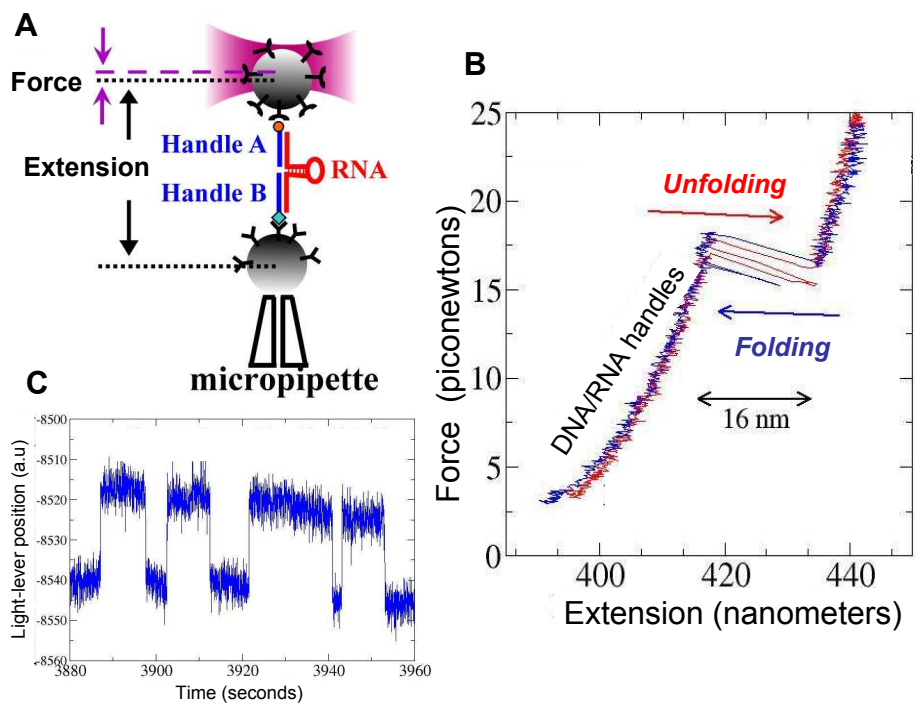}
  \caption{RNA pulling experiments using LOT. (A) Experimental setup. An
  RNA molecule is inserted between two hybrid RNA/DNA handles. The force
  (F) and extension (X) of the molecular assembly are measured as the
  micropipette is moved. (B) FEC of a 20 bps RNA hairpin in a pulling
  cycle. During the stretching part of the cycle (red curve) the elastic
  response of the handles is followed by the sudden release of extension
  and a drop in the force corresponding to the unfolding of the RNA
  molecule. During the relaxation part of the cycle (blue curve) the RNA
  molecule folds back again. (C) Hopping experiments. If the force is
  held constant the RNA molecule hops between the folded and unfolded
  conformations. The vertical axis represents the position of a reference
  laser beam for the chamber which is related to the molecular extension.}  \label{fig6}
\end{figure}

Glass microfibers, AFM and LOT have been used to stretch and unzip RNA
molecules. The latter is by now the most accurate technique to resolve
forces on the order of the pN and distances on the order of a few
nanometers that are observed in the unfolding of small RNA hairpins. The
first RNA pulling experiments were carried out in the Bustamante lab
\cite{LipOnoSmiTinBus01} where the P5ab RNA hairpin (a derivative of the
L121 Tetrahymena rybozyme) was studied.  In such experiments an RNA
hairpin is stretched in a force-ramping experiment and the FEC
recorded. The molecule is found to unzip at forces around 15pN resulting
into a denaturated single stranded RNA form (ssRNA). The RNA molecule
also hops between the folded and the unfolded states at a critical value
of the extension or force applied on the molecule where the folded and
unfolded conformations are equally populated. Pulling RNA (or DNA)
hairpins has additional complications as compared to the case of
stretching dsDNA. Being the hairpin a few tens of base pairs long the
gain in extension after the molecule unfolds is between 10 and
50 nm. Such extension is too short to manipulate the RNA molecule using
micron-sized beads. To pull on the RNA hairpin, two hybrid RNA/DNA
handles are synthesized and annealed to the RNA molecule at its flanking
sides. The handles are typically a few hundreds base pairs long so the
whole construct is approximately a few hundred nanometers long when fully
extended. The experimental setup and a typical FEC is shown in
Fig.~\ref{fig6} (A-B). Upon increasing of the force the RNA hairpin
unfolds, the extension of the molecule increases and the force drops in
response to the retraction of the trapped bead that follows to the
sudden gain in molecular extension. If the force is relaxed back then
the hairpin refolds again showing a sudden increase in the force upon
formation of the hairpin. Several
thermodynamic and kinetic properties can be investigated in these
experiments. By pulling slowly enough (the lowest value of the pulling
speed being limited by low-frequency drift effects in the instrument)
one can infer the mechanical work necessary to unfold the hairpin which
is equal to the area below the FEC.  After subtraction of the reversible
work required to stretch the hybrids handles and the unfolded ssRNA, the
free energy difference between the folded and unfolded state at zero
force and room temperature can be inferred \cite{LipOnoSmiTinBus01}.
The value of the folding free energy obtained for RNA secondary
structures agrees well with theoretical estimates by Mfold
\cite{Zuker89,MatSabZukTur99}. By repeatedly pulling-relaxing the
molecule many times, hysteresis is often observed in the FECs. The
average unfolding force is always larger than the average refolding
force. Kinetic properties can be studied by pulling the hairpin at
different rates and measuring the breakage or dissociation force
distribution during unfolding.  Combined with the measurement of the
refolding force distribution along the retracting part of the cycle it
is possible to identify the location of the transition state and the
free-energy landscape of the molecule as a function of its extension
\cite{ManColRit05}. Pulling experiments using LOT are usually carried
out near equilibrium conditions. AFMs allow to perform dynamic force
spectroscopy measurements of RNA dissociation far from equilibrium by
exploring a few orders of magnitude of loading rates
\cite{GreWilWahDavRobTenAll04}. These studies reveal that the average
dissociation force increases logarithmically with the pulling speed as
has been found in the study of intermolecular protein-protein
interactions (see Sec.~\ref{proteinprotein}).  Similar experiments have
been carried out to investigate the kinetics of short DNA hairpins using
SMF \cite{BonKriLib98,AltLibKri03} or AFM
\cite{LeeChrCol94,NoyVezKayMeaLie97,StrOroSchGun99,GraStrSchGun01,SchGraStrBerGunHeg02}
finding slower kinetics of unfolding/refolding depending on the length
of the sequence as predicted by some theoretical models
\cite{LubNel02,Nelson03}. Mechanical unfolding of single RNA molecules
through nanopores has been also proposed as a method to determine the
secondary structure \cite{GerBunHwa04}. The unzipping of RNA hairpins
has motivated several theoretical studies of thermodynamic
\cite{MonMez01,GerBunHwa01,DesMaiZhaPelBenCro02} and kinetic properties
\cite{ImpPel04,HyeThi05,LiuYan04,LiuTonYan05,HyeThi06}.

Other related experiments provide additional insight on the kinetics of
unfolding of the molecule. For example, if a constant force that is
close to a critical value is maintained by a force feedback mechanism
then the RNA molecule hops between the folded and unfolded states
(Fig.~\ref{fig6}C) (the mechanism being similar to the opening and
closure activity that is observed in single ion channels
\cite{FerChuOber01}). Hopping can be investigated in two different
modes: 1) the passive mode where the position of the micropipette and
the trap are kept fixed; 2) the force-feedback mode where the force is
maintained constant by using a piezo controller that corrects the
position of the micropipette every time there is a change in the force
(see Sec.~\ref{lot}). Working at the vicinity of the critical extension
or force (where the molecule hops between the folded and unfolded
conformations) it is found that the molecule follows exponential
kinetics. The probability distribution of the residence times in the
folded (unfolded) conformations is well described by an exponential
function whose width is equal to the inverse of the kinetic rates of
unfolding (folding). The dependence of these rates on the applied force
gives accurate information about the height and position of the kinetic
barrier.  The hopping kinetics of RNA hairpins has been modeled by Cocco
and collaborators who have introduced a one-dimensional representation
of the possible configurations of the molecule in terms of the number of
sequentially open base pairs starting from the beginning of the fork
\cite{CocMonMar03,GerBunHwa03}. Models for the experimental setup where
the distance between the micropipette and the center of the trap are the
appropriate control parameter have been considered in \cite{ManRit05}
whereas a detailed analysis of the influence of the experimental setup
(length of the handles, stiffness of the trap, bandwith of data
collection, time delay of the force-feedback mechanism) on the
measurement of the intrinsic molecular kinetic rates of the RNA molecule
in the different modes (passive or force-feedback) has been carried out
in \cite{Man05a,Man05b}. Passive force clamp methods operating without
force-feedback have been also implemented in dual laser traps by taking
advantage of the anharmonic region of the trapping potential. Studies of
the hopping kinetics of a tetraloop DNA hairpin show that the hopping
frequency increases with the stiffness of the trap
\cite{GreWooAbbBlo05,WooBehLarTraHer06}.

Two-states behavior is usually found during the unfolding and
refolding of short RNA hairpins
\cite{LipOnoSmiTinBus01,RitBusTin02}. Experiments have been carried
out in more complex molecules such as the full {\em Tetrahymena
thermophila} rybozyme L21 (containing approximately 400 nucleotides)
using FRET \cite{ZhuBarBabRusHaHerChu00} or LOT
\cite{OnoDumLipSmiTinBus03}. LOT made also possible the mechanical
unfolding of the {\em Escherichia coli} 1540 bps-long 16S ribosomal
RNA by the group of D. Chatenay \cite{HarMarRobLegXayIsaCha03}. The
resulting FECs reveal a series of complex rips that correspond to the
opening of different domains of the molecule. Other studies have
investigated specific RNA motifs such as internal bulges
\cite{HarMarRobLegXayIsaCha03} , the transactivation response region
(TAR) RNA derived from the human immunodeficiency virus (HIV)
\cite{LiColSmiBusTin06} and three-helix junction RNA molecules
\cite{HaZhuKimOrrWilChu99,KimNieHaOrrWilChu02,ColRitJarSmiTinBus05}.
These experiments show that force unfolding proceeds through the
successive opening of domains whereas force refolding is a much more
complex process where various folding pathways and trapped
intermediates are often observed
\cite{ZarWil94,PanThiWoo97,TreWil99,TreWil01}. The kinetic behavior of
RNA molecules shares many resemblances to what has been observed in
proteins (see Sec.~\ref{folding}), often showing misfolded structures
\cite{LiBusTin05}, reinforcing the observation that the free-energy
landscape underlying the folding dynamics is more rugged in RNA than
in proteins \cite{CheDil00,HyeThi03,ThiHye05}.


\subsubsection{DNA condensation.}
\label{condensation}

The nuclear genome is not isolated but surrounded by many different
proteins engaged in its maintenance and regulation
\cite{Watson04,CalDre97}.  Proteins are also responsible for the
compaction of eukaryotic DNA inside the nucleus of the cell
\cite{Bloomfield96}. The nuclear DNA is condensed with proteins into a
huge molecular complex called chromatin. Linear compaction defined as
the ratio between the length of fully extended DNA to the length of
the condensed DNA reaches values on the order of $10^4-10^5$.  The basic
unit of the condensed DNA is the nucleosome core particle, a flat disk
of 11 nm diameter with 146 bps of supercoiled DNA wrapped around a
histone octamer formed by pairs of histones H2A, H2B, H3, H4. The main
force stabilizing the nucleosome particle is the electrostatic
attraction between the negatively charged phosphate backbone of DNA
and the protonated (positively charged) arginine and lysine lateral chains of
histones.  Nucleosome particles are destabilized and readily
dissociate at low salt concentrations \cite{YagMurHol89}. Nucleosomes
are connected to each other by segments of variable length of linker
DNA (around 60 bps) forming a {\em beads in a string} structure.  The
histone H1 stabilizes the structure of the nucleosome by fixing the
entry and exit angle of the wrapping DNA. Chromatin organizes into
different structures at different lengthscales \cite{Hansen02}. The
nucleosome is the minimal unit in such organization, also called the
10 nm fiber. At physiological salt values nucleosomes form a complex
and dense structure recognized as the 30 nm fiber formed by the folding
of nucleosomes into a yet unknown three-dimensional structure
\cite{Widom98,Daban03,Schiessel03}.

\begin{figure}
\centering   \includegraphics[scale=.55,angle=-90]{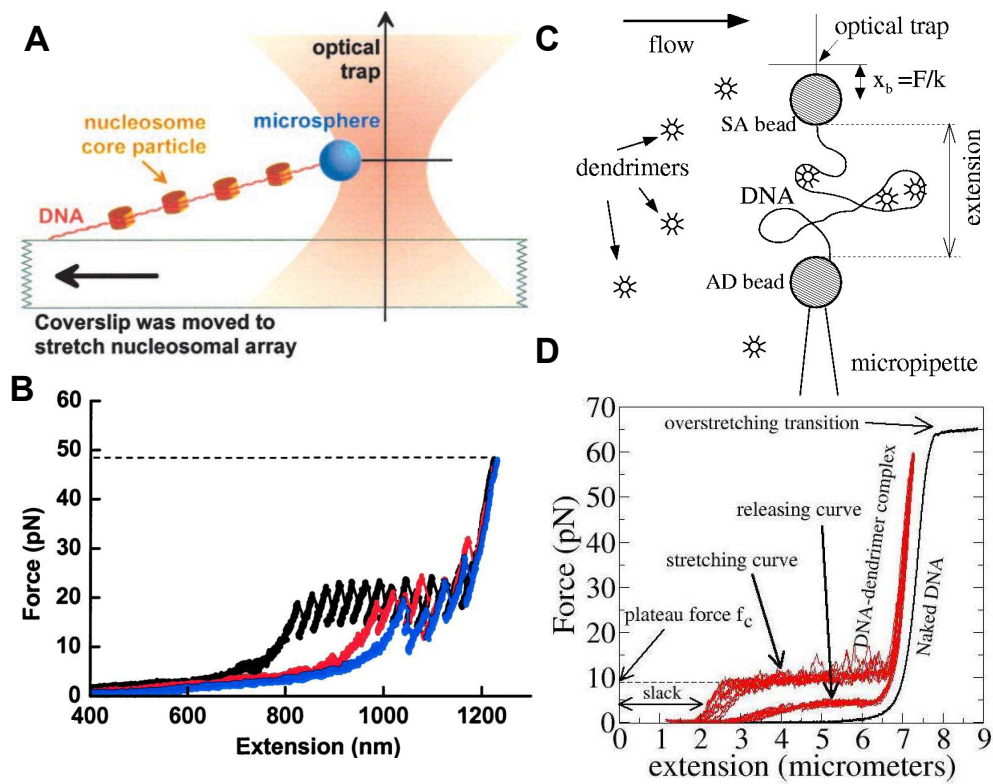}
  \caption{FECs in DNA condensation using LOT. (A) Mechanical
disruption of nucleosomal arrays. Experimental setup. (B) Nucleosomal
arrays repeatedly stretched three times (first -black-, second -red-,
third -blue-) up to a maximum force of 50pN. The unwrapping of
individual nucleosomes is observed as jumps in the FEC. As the array
is repeatedly pulled the number of remaining nucleosomes in the fiber
decreases. Figure taken from \cite{BroSmiYehLisPetWan02}. (C)
Condensation of DNA molecules with polyaminoamide (PAMAM) dendrimers
(Experimental setup). (D) FECs of DNA fibers (red curves) condensed
with dendrimers of generation G8 (diameter of particles around 10 nm)
\cite{RitMihSmiBus05}. The black curve is the FEC of naked DNA before
condensation. The red curves show hysteresis between the stretch and
release part of the cycle, the presence of a slack (due to the
presence of uncondensed segments in the fiber) and a decondensing
force plateau around 10 pN.}
\label{fig7}
\end{figure}

There are few experimental single-molecule studies of DNA
condensation. The first study used LOT to pull chicken erythrocyte
chromatin fibers \cite{CuiBus00}. It showed irreversible force-extension
cycles above 20 pN interpreted as due to the mechanical removal of
histone cores from native chromatin. SMF and AFM imaging have explored
the assembly kinetics of chromatin from Xenopus eggs and Drosophila
embryos \cite{LadQuiDoyRouAlmVio00}. None of these preliminary studies
could identify the dynamics of individual nucleosomes. Subsequent
studies analyzed the condensation of reconstituted chromatin fibers of
$\lambda$-phage DNA suspended between two beads and exposed to {\em
Xenopus laevis} egg extract
\cite{BenLeuLenZlaGroGre01,BenPopZlaGroGre01}. The fiber condensed under
a constant frictional Stokes force. The rate of condensation decreased
considerably under applied force and no condensation events were
observed at forces exceeding 10 pN. Strong inhibition of chromatin
assembly in chromatin fibers above 10 pN has been also observed in MT
studies \cite{LeuKarTomRamSmiZla03}. After the fiber had condensed
subsequent FECs revealed a series of force jumps between 15 and 40 pN
attributed to the release of the wrapped histone octamer. After all
releasing events have taken place the FEC characteristic of naked DNA
was recovered. However, due to the large amount of different type of
proteins in the cell extract, it is difficult to tell which jumps
correspond to other bound proteins and which jumps are due to
the unravelling of the histone octamer. Similar experiments have been
carried out in the study of nucleosomal arrays of reconstituted pure
histones \cite{BroSmiYehLisPetWan02}. FECs reveal a series of disruptive
events attributed to the unwrapping of individual nucleosomes containing
80 bp of dsDNA (Fig.~\ref{fig7}A) . Single-molecule force measurements to
test higher order chromatin structures have been applied to investigate
the protein scaffold of mitotic chromosomes \cite{PoiMar02}.

Recent studies of DNA condensation have considered synthetic condensing
agents much simpler than protein histones such as shell crosslinked
nanospheres \cite{ThuRemKowWoo99} and dendrimers
\cite{KukBieJohSpiTomBak96}. Dendrimers are synthetic branched polymers
that are synthesized via an initiator core terminating (after a repeated
series of steps) with amino $NH_2$ groups on the surface
\cite{FreTom01,Frechet02}. Questions such as the importance of charge,
shape and size in the condensed state have been recently considered by
studying PAMAM (polyaminoamide) dendrimers condensed with
$\lambda$-phage DNA using LOT \cite{RitMihSmiBus05}. FECs in dendrimers
show a force plateau around 10pN, characteristic of a decondensation
transition between a condensed and an extended phase (Fig.~\ref{fig7}B).
These experiments also reveal that residual inter-dendrimer
electrostatic interactions keep the structure of the condensed state as
revealed by the salt dependence of the force value from the {\em
plateaux}, similar to what has been found in the condensation of DNA by
other polycationic complexes such as eukaryotic condensin and trivalent
cations like hexaammine cobalt (CoHex) and spermidine
\cite{BauSmiBloBus97,DenBlo99,BauBloSmiBusWanBlo00}. The mechanism by
which nucleosomes form and arrange into a compact globular structure has
been the subject of many theoretical and numerical studies
\cite{WooGriHorWhi93,KunNet00,KatBusOls00,NguGroShk01,NguShk01,SchWidBruGel01,SchGelBru01,GroNguShk02,KunNet02,BorNet03,KulSch04,WadMurSan05,SunZhaSch05}.

\subsection{Proteins}
Many single-molecule studies have investigated in detail the
mechanochemistry of some relevant proteins. Two category of protein
systems must be distinguished: 1) Those where an energy source (e.g.
from ATP or GTP hydrolisis) is not required; 2) those where catalytic
functions driven by high energy phosphate compounds (e.g. ATP or GTP
hydrolisis) are necessary.  The first class of proteins is closer in
spirit to manipulation experiments of single nucleic acids in which
molecular interactions are probed either by mechanical force or observed
using optical imaging techniques. The second class of
proteins includes the study of many molecular motors and enzymes. Due
to the large number and the importance of the studies on ATP-dependent
motors a whole section is devoted below to them.

\subsubsection{Protein-Protein interactions.}
\label{proteinprotein}

Proteins have important regulatory functions as demonstrated by Jacob
and Monod in the early 60's with the discovery of the Lac
repressor. Proteins can participate in chemical signaling at the intra
and intermolecular level. The successful development of an organism
relies on the coordination of myriads of signals transmitted among a
large number of proteins all participating in a structurally complex and
highly dynamic web of interactions.  Developmental processes such as
cellular association, differentiation, patterning and reproduction all
are controlled mainly by proteins.

\begin{figure}
\centering \includegraphics[scale=.55,angle=-90]{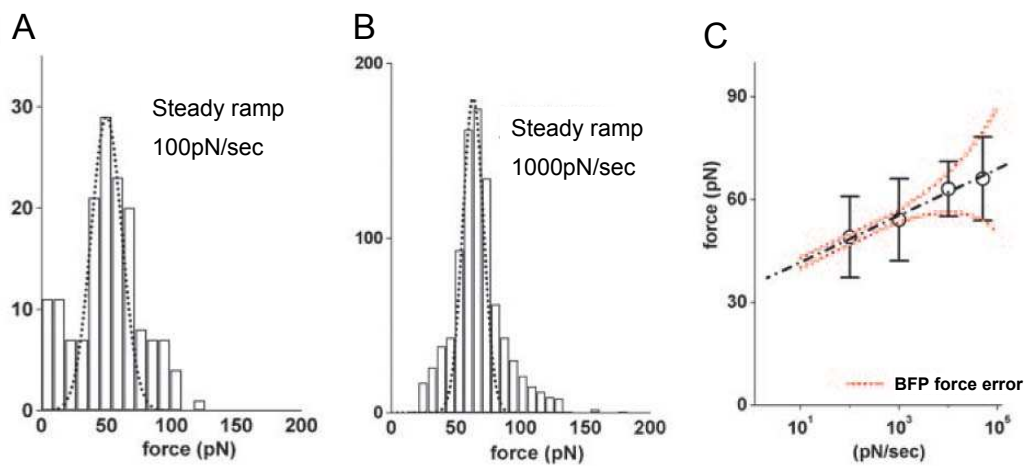}
  \caption{Mechanical strength of trans-bonded cadherin fragments. (A,B)
  Force histograms in steady ramp protocols at 100 pN/sec and
  1000 pN/sec. Force distributions shift to larger forces as the loading
  rate increases (C) Logarithmic rate dependence of the average rupture
  force over nearly 4 orders of magnitude in loading rates. The dotted
  line corresponds to the measurement error in force. The error
  increases with the loading rate because of viscous corrections
  appearing due to probe damping. Figure taken from
  \cite{PerLeuFerEva04}.}
\label{fig8}
\end{figure}

A large of number of studies have focused in the study of
protein-protein interactions of the ligand-receptor type
\cite{RosEckBarSisBauWilPuhSewBecAns04}. These interactions are governed
by what in biological terms has become known as "lock and key"
mechanism. Proteins can interact and {\em recognize} each other by
assembling into larger complexes by mutual complementarity and fit of
their molecular surfaces. Allostery refers to the regulation of such
interaction by changes in either one or both proteins. Protein-protein
interactions are studied in dynamic force spectroscopy using AFM, BFP
and LOT. The first type of measurements are the most common. Samples are
prepared by coating a substrate with ligand molecules and the tip of an
AFM with receptor molecules. The substrate is then moved towards the tip
until a few contacts between ligand and receptor molecules are
formed. Ideally one would like to have just one molecular contact
between the tip and the substrate. In general this is not possible and
the concentration of the proteins has to be carefully tuned to ensure
that just one connection is often established between tip and substrate
(implying that the large majority of contact trials are unsuccessful and
a connection is rarely established).  Upon retraction of the tip the
extension of the protein-protein contact increases and the force
increases up to a value where the contact breaks. The rupture of the
contact is stochastic, therefore upon repetition of the experiment
several times (each time a new contact has to be sought) the value of
the rupture force is always different, the overall rupture process being
described by a breakage force distribution. The rupture force
distribution depends in a non-trivial way on the retraction rate of the
tip. Typically, the larger is the retraction rate the higher is the
average rupture force which grows approximately as the logarithm of the
loading rate. Kramer theories for chemical reactions
\cite{Kramers40,HanTalBor90,Melnikov91} extended by Bell in the 70's to
include the effect of mechanical force on bond
dissociation\cite{Bell78}, have been adapted by Evans and Ritchie
\cite{EvaRit97,EvaRit99,Evans01} to interpret the observed rate
dependencies. Dynamic force spectroscopy
\cite{MerNasLeuRitEva99,EvaWil00} has become nowadays a standard method
to probe the strength of molecular bonds. It has been applied to the
study of biotin-avidin and biotin-streptavidin interactions
\cite{LeeKidCol93,FloMoyGau94,MoyFloGau94}, antigen-antibody
interactions
\cite{DamHegAnsWagDreHubGun96,HinBaumGruSchSch96,SchRosStrAnsGunHonJerTiePlu00},
P-selectin/ligand interactions
\cite{FriKatKolAns98,HanCarJadTseWirKon03}, adhesion forces in lipid
bilayers \cite{GarOncSan05}, substrate-protein adsorption forces
\cite{GerVoeSchSenMaaHorHem00}, cadherin mediated intermolecular
interactions \cite{PerBenNasPieDelThiBonFer02,PerLeuFerEva04} ,
carbohydrate-protein bonds \cite{EvaLeuSim01}, proteoglycans
\cite{DamPopWagAnsDreHubGun95,GarBucRosAnsSanBurFer06}, antibody-peptide
interactions \cite{SulFridLanLauAlbRatDeNColNoy05} just to cite a few
examples. Typical rupture force distributions are shown in
Fig.~\ref{fig8}. The problem of bond rupture has been extended to
multiple bonds in several configurations (series, parallel, zipper)
\cite{EvaWil00,BarDerAdj02,Williams03}. An annoyance inherent to dynamic
force spectroscopy in ligand-receptor studies is that after a rupture
event occurs a contact has to be established again. As discussed below,
this is different from what happens when studying intramolecular
interactions where the initial set of bonds can be reformed again by
moving back the surface to the tip.

\subsubsection{Protein folding.}
\label{folding}

\begin{figure}
\centering   \includegraphics[scale=.6,angle=-90]{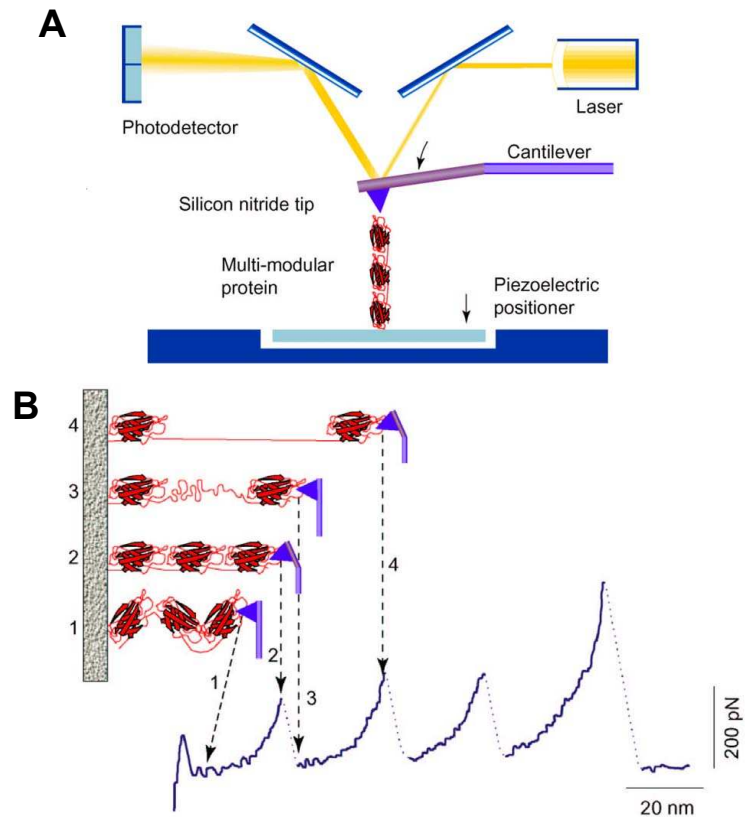}
  \caption{Pulling multi-modular proteins with AFM. (A) Experimental setup. A
  laser beam is deflected by the cantilever tip of the AFM. The amount of the
  deflection is detected in a position sensitive detector (PSD). (B) The
  characteristic {\em saw-tooth pattern} of force obtained by pulling on a
  recombinant construct of Ig27 domains from titin. Figure taken from  
  \cite{FisObeCarMarFer99}.}  
\label{fig9}
\end{figure}

SME allow to investigate intramolecular interactions in
proteins. Pioneering studies have been carried out by pulling the
muscle protein titin. The sarcomere is a repeating unit found in
fibers of muscle cells responsible for contractile motion. The
sarcomere is a highly complex structure 2.5 $\mu$m long made out of
thick (myosin) and thin (actin) filaments and a {\em third filament}
(titin). Titin connects the Z-disk to the M-line in the sarcomere, and
is a huge modular protein responsible for the passive elastic
properties of the muscle. Titin is formed by tandem pseudorepeats of
many different protein domains such as the immunoglobule (Ig) and the
fibronectin type III (fnIII) domains. The characterization of the
mechanical properties of modular proteins like this one is very
important because they are present in the cytoskeleton and the
extracellular matrix of all eukaryotic cells.  AFM and LOT pulling
experiments in proteins were first carried out in titin
\cite{RieGauOesFerGau97,KelSmiGraBus97,RieGauSchGau98} by the groups
of H. Gaub in Munich and C. Bustamante in Berkeley. Upon application
of force titin unravels in a series of force jumps, one jump
corresponding to the unfolding of an individual module. A FEC reveals
a characteristic sawtooth pattern as shown in Fig.~\ref{fig9}. Due to
the heterogeneous structure of the module sequence in modular proteins
(such us titin) it is difficult to identify which specific module
corresponds to which unfolding event. Considerable progress has been
later achieved in the group of J. Fernandez using protein engineering
techniques to construct homomeric polyproteins, i.e. tandems of
identical repeats of a protein such as the I27 module of titin
\cite{CarObeFowMarBroClaFer99,CarMarObeFer99,CarObeFisMarLiFer00}
(similar constructs have been synthesized in the T4 lysozyme
\cite{YanCecBaaBetBreHaaMatDahBus00}). The study of polyproteins of
I27 allows to carry out detailed studies of the mechanical stability
and unfolding/refolding kinetics of a single molecule.

Similar studies have been conducted in other modular proteins such as
tenascin \cite{ObeMarEriFer98}, triple helical coiled coils of spectrin
\cite{RiePasSarGau99} bacteriorhodopsin \cite{MulKesOesMolOesGau02} and
the cellular adhesion molecule Mel-CAM \cite{CarKwoManSpeDis01}.  In all
cases the unfolding kinetics is well described by a two-state process
where the distribution of ripping forces and FECs depends on the loading
rate. The distributions are also influenced by the polymer spacers and
the instrument limitations \cite{RieFerGau98,FriWehKuhGau03}.  These
studies provide insight also on the role of the different structural
elements of the protein on its mechanical stability, such as
$\beta$-sheets and $\alpha$-helices, the main building blocks of the
secondary structure in proteins. Experimentally it has been also found
that as a rule $\beta$-sheets are more robust elements than
$\alpha$-helices \cite{MunThoHofEat97}. The geometry of the pulling,
whether unzipping or shearing $\beta$-strands, and the point of
application of the force also determines the final mechanical stability of
the protein \cite{CarObeFisMarLiFer00,Car05}. The influence of specific
solvent conditions on protein stability can also be studied using the
AFM. In this line interesting studies have been carried out by the
Discher's group on the mechanical stability of a vascular cell adhesion
molecule (VCAM-1), a tandem of seven Ig domains that are stabilized by
disulfide bonds and that can be destabilized in the presence of reducing
agents \cite{BhaCarHarFenLuSpeDis04}. Simultaneous force experiments and reduction of disulfide bonds
(to SH) can be studied on single molecules
\cite{BhaCarHarFenLuSpeDis04,WiiAinHuaFer06,DisBhaJon06}. Finally, the study of force
unfolding kinetics in proteins has motivated various theoretical and
numerical studies \cite{KliThi99,PacKar00,SheCanCam02}.

Protein folding, the process by which a molecule folds into a three
dimensional functionally active structure, is still a major unsolved
problem in modern biophysics. The dynamics of such process has many
aspects and complications, for example many proteins need to be assisted
by some other proteins (chaperons) to become efficiently and quickly
folded \cite{ValMarGomVarWil02}. A fascinating aspect of SME is the
possibility to investigate folding under applied force in a single
molecule, i.e. the equivalent of protein folding in bulk experiments but
with force being the externally controlled variable. Upon pulling the
protein is unfolded, while upon retraction of the positioner the protein
can fold back again into its native structure. The process should be
similar to that observed in RNA molecules. However, there are two
inconveniences or difficulties that make the study of force-folding of
proteins with AFM more difficult than RNA folding using LOT. The first
difficulty is found in the high value of the stiffness of the AFM tip
(typically 100 times larger than that of a LOT, see Sec.~\ref{afm})
which considerably increases thermal fluctuations in the force making
difficult to control the value of the force during the refolding
process. The second inconvenience is related to the experimental
difficulties of unequivocally identifying individual molecules due to
many non-specific interactions between tip and substrate. To overcome
this second limitation the synthesis of polyproteins has been used in
order to identify true unfolding events. The study of force-folding
kinetics in proteins has been carried out by measuring the force-clamp
relaxation (i.e. the force is maintained constant by using a feedback
loop) of polyproteins that have been previously mechanically
unfolded. However, it is unclear whether in such conditions the folding
kinetics of individual modules might be affected by the folding kinetics
of other modules. Recent AFM measurements by the group of
J. Fernandez in polyubiquitin proteins (ubiquitin is an ubiquitous and
highly conserved eukaryotic protein in charge of signalling for the
proteosomal degradation of proteins, among other functions) using the
force-clamp technique \cite{SchLiFer04} have shown that the refolding
kinetics is not a two-state process but a multiple pathway dynamical
process determined by the roughness of the free-energy landscape of the
molecule \cite{FerLi04,BruHerWalFer06} (see however
\cite{Sosnick04,FerLiBru04,BesHum05,FerLi05a} for a discussion of these
results and alternative interpretations).  Similar observations have
been reported in RNA pulling experiments \cite{LiBusTin05}. These
studies suggest the existence of intermediate states along the
unfolding/folding pathway of proteins. Evidence for intermediates has
been obtained in the study of the unfolding kinetics of titin with AFM
\cite{LuIsrKraVogSch98,MarLuLiCarObeSchFer99,WilFowBesTocScoSteCla03}.
Recently, LOT have been also used to investigate the folding kinetics of
RNAseH enzyme revealing the presence of an intermediate conformation of
the protein, probably the molten globule state observed in thermal
denaturation experiments \cite{CecShaBusMar05}. This work paves the way
to investigate other proteins using LOT. SME in proteins are expected to
greatly contribute to our current understanding of protein folding from
a statistical physics approach \cite{OnuLutWol97,Finkelstein03,JunRit06}.

Fluorescence spectroscopy has been also used to investigate and detect the
presence of intermediate states in proteins. Different methods are used to
immobilize and observe proteins: passive adsorption, specific tethering,
trapping inside polymer gels or vesicle encapsulation \cite{Haran03}. In this last
technique proteins are encapsulated inside vesicle cells of size larger
than the protein but smaller than the laser beam section allowing for a
mild immobilization of the protein. Haran and collaborators have shown
that two-state folders, well characterized by bulk biochemistry methods,
display exponential kinetics in their relaxation \cite{RhoCohSchHar04}. Adenylate kinase, an
enzyme 214 amino acids long that catalyzes the production of ATP from ADP, shows a
heterogeneous and slow dynamics characterized by multiple folding
pathways \cite{RhoGusHar03}. These results are in agreement with the conclusions obtained
in the aforementioned force-clamp AFM studies on polyubiquitin \cite{FerLi04}
and FRET studies on RNAse P (a transfer RNA that contains a catalytic RNA
subunit) \cite{XieSriSosPanSch04}. 

\subsection{Molecular motors}
\label{motors}

\begin{figure}
\centering \includegraphics[scale=.45,angle=-90]{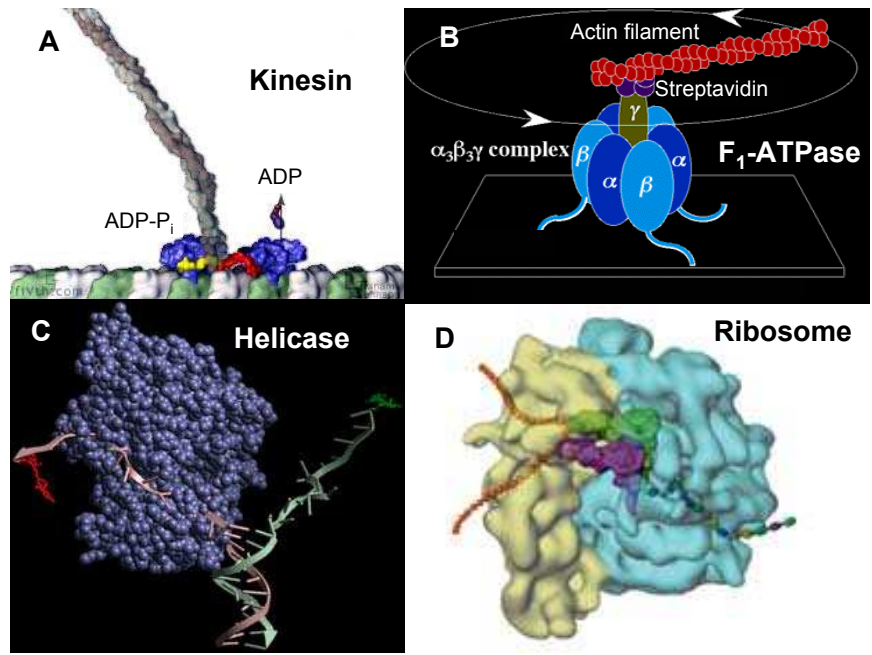}
  \caption{Examples of molecular machines: (A) Kinesin walking along a
  microtubule and transporting a cargo (not shown). (B) $F_1$-ATP synthase is the
  proton pump responsible of synthesizing ATP in the mitochondria of
  eukaryotic cells. (C) Helicases are forerunners of the DNA polymerase
  that unwind DNA by transforming dsDNA into two strands of ssDNA. (D)
  The ribosome is among the largest molecular machines inside the
  cytoplasm of the cell and is in charge of manufacturing proteins.}
\label{fig11}
\end{figure}

One of the great applications of SME is the possibility to follow
individual molecular machines in real time while they carry out specific
molecular tasks \cite{Spudich02}. Molecular motors are proteins that
typically use the energy extracted from available sources, such as
chemical gradients or high energy phosphate compounds (e.g. ATP or GTP)
to exert mechanical work (Fig.~\ref{fig11}). For example, in ATP
hydrolisis a molecule of ATP is broken into ADP and inorganic phosphate
$P_i$ (${\rm ATP}\to {\rm ADP}+{\rm P}_i$) in a highly irreversible reaction that releases
around $7~{\rm kcal/mol}\simeq 12~k_BT$. The chemical process by which
motors utilize the energy stored in the high energy bonds of the ATP
molecules to perform mechanical work is based on two hypothesized
mechanisms: 1) Power stroke generation or 2) Brownian ratchet
mechanism. In the first mechanism either the production of ADP or the
release of the inorganic phosphate during the ATP hydrolysis cycle
induces a conformational change in the substrate that is tightly coupled
to the generation of force and motion in the motor. In the second
mechanism, the motor diffuses reversibly along the
substrate. Unidirectional movement is produced by rectification of
thermal fluctuations induced by the conformational change in the protein
caused by ATP hydrolisis. By steady repetition of a mechanochemical
cycle (one or more ATP molecules are hydrolyzed per cycle) the motor
carries out important cellular functions. Motors are characterized by
the so called processivity or number of turnover cycles the motor does
before detaching from the substrate. Processivities of molecular motors
can vary by several orders of magnitude, depending on the type of motor
and the presence of other regulating factors. For example, the muscle
myosin II motors work in large assemblies, each myosin having a
processivity around 1, meaning that each myosin performs one
mechanochemical cycle on average before detaching from the substrate. In
the other extreme of the scale there are DNA polymerases (DNApols,
enzymes involved in the replication of the DNA) in eukaryotes which show
processivities that range from 1 (adding approximately one nucleotide
before detaching) up to a several thousands or even millions. However,
in the presence of sliding clamps (proteins with the shape of a
doughnut that encircle the DNA and tightly bind DNApols
\cite{JerDonKur02}) processivities go up to $10^9$. 

Molecular motors are
magnificent objects from the point of view of their efficiency. If we
define the efficiency rate as the ratio between the useful work
performed by the motor and the energy released in the hydrolysis of one
ATP molecule in one mechanochemical cycle, then typical values for the
efficiencies are around several tens per cent reaching also the value of
97$\%$ in some cases (like in the $F_1$-ATPase, see below). For example,
out of the 20 $k_BT$ obtained from ATP hydrolysis,
kinesin can exert a mechanical work of 12 $k_BT$ at every step, having
an efficiency of around 60$\%$. Such large efficiencies are rarely found
in macroscopic systems (motors of cars have efficiencies in the range
20-30\%) meaning that molecular motors have been designed by evolution
to efficiently operate in a highly noisy environment
\cite{BusKelOst01}. Molecular motors are expected to be essential
constituents of future nanodevices. Single-molecule devices that operate
out of equilibrium and are capable of transforming externally supplied
energy into mechanical work are currently being studied
\cite{HugHolCatMorSeiGau02}.

Two major classes of molecular motors have been experimentally studied
using single-molecule techniques. One class corresponds to molecular
motors involved in several transport processes inside the cell cytoplasm
such as the aforementioned kinesin, dynein and myosin
\cite{Howard01}. Single-molecule measurements in molecular motors were
carried out by Block and coworkers, who studied a single kinesin on a
microtubule rail. Kinesin and dynein are both microtubule associated
proteins (MAPs) involved in the transport of organelles and vesicles
along microtubules in the cytoplasm of the cell. They move in opposite
directions with respect to the polarity of the microtubule. Using LOT it
is possible to attach one kinesin molecule to a bead captured in an
optical trap and measure the force exerted on the bead while the kinesin
walks along the microtubule \cite{Cross04,CarCro06}.  Kinesins are found
to move in 8 nm steps at an average speed of 2 microns per second
\cite{SvoSchSchBlo93}. Kinesins move in a hand-over-hand way by
alternate exchange between the head of the molecule (that makes a strong
bond with the $\beta$-tubulin domain of the microtubule while ATP is
hydrolyzed) and the other head that detaches from the substrate
\cite{YilTomValSel04}. The average velocity of kinesin has been shown to
depend on the value and direction of the applied force
\cite{UemIsh03,BloAsbShaLan03} with a stall force around 7 pN required
to arrest the motion \cite{BloGolSch90,CarCro05}. Kinesin also shows
backward steps during its motion suggesting a Brownian ratchet mechanism
\cite{NisHigYan02,CarCro05}. Depolymerazing kinesin motors of the
cytoskeleton and extraction of membrane nanotubes by kinesins have been
also investigated in SMF experiments
\cite{HelBroKalDieHow06,LedCamZelRouJolBouGouJoaBasPro04}.  Dynein has
also been studied and shown to display steps that are multiple of 8 nm
\cite{MalCarLexKinGro04} depending on the applied force. The force
generation mechanism is still unknown \cite{BurWalSakKniOiw03}.

Myosins are a large class of proteins responsible for muscular cell
contraction that walk along actin filaments \cite{Sellers99}. Two main
types of myosins have been studied in SME: monomeric (single head) and
dimeric (two heads) motors. Examples of single head myosins are
myosin-I, myosin II and myosin IXb
\cite{VeiColJonSpaMilMol99,KitTokIwaYan99,InoSaiIkeIke03}. Examples of
two-headed myosins are myosin V and myosin VI
\cite{VeiWanBarSelMol01,TanHomIwaKatIkeSaiYanIke02,ForQuiShaCorGol03}. LOT
and SMF measurements have shown that myosins moves along actin in steps
of average size about 10 nm in single head myosins
\cite{FinSimSpu94,MolBurKenTreWhi95} and 36 nm for the two headed myosin
V \cite{YilForMcKHaGolSel03}.

Another motor that has been extensively studied
using SMF techniques is the protein machine $F_0,F_1$-ATP synthase, a
proton pump that synthesizes ATP in the mitochondria, the {\em power plant} of
eukaryotic cells. The $F_0,F_1$-ATP synthase dissociates in two parts,
$F_0$ (the proton channel) and $F_1$ (often called $F_1$-ATPase). The
latter is a complex made out of a shaft containing two different types
of 3 tubular subunits each (called $\alpha$ and $\beta$) and a central
subunit called $\gamma$. By engineering appropriate mutations in the complex,
Kinosita and coworkers \cite{NojYasYosKin97} immobilized the shaft to a
surface and attached the central $\gamma$ subunit to a fluorescent actin filament
(Fig.~\ref{fig11}B). In the presence of ATP the actin filament was
observed to rotate in steps of 120 degrees
\cite{YasNojKinYos98,AdaYasNojItoHarYosKin00,MasMitNojMunYasKinYos00}
which were later resolved into substeps of 30 and 90 degrees, each
substep corresponding to different phases of the mechanochemical cycle
of ATP synthesis \cite{YasNojYosKinIto01}. $F_1$-ATPase motors have
also the capability to be integrated with nanoelectrochemical systems
to produce functional nanomechanical devices
\cite{SooBacNevOlkCraMon00}.

A second class of motors are those that interact with the DNA and
participate in maintenance tasks of the genome such as transcription
and replication. Examples are DNA polymerases (DNApols), DNA
translocases, RNA polymerases (RNApols) and DNA
topoisomerases. DNApols have been studied using LOT and MT. Studies
have been carried out in \cite{WuiSmiYouKelBus00}
with the T7 DNApol (belonging to the bacteriophage T7 and
characterized by having a high processivity of several thousands base
pairs) and in the Bensimon group \cite{MaiBenCro00} who studied the
sequenase (a mutant of T7 DNApol lacking the exonuclease site) and the
Klenow DNApol (a fragment of {\em Escherichia coli} DNApol I lacking
exonuclease activity). In these experiments a ssDNA containing a DNA
primer required for the initiation of replication is tethered between
one immobilized surface and one detector bead that measures the force
acting on the molecule. After flowing DNApol, ATP and other
nucleotides (NTP) inside the chamber the tether extension of the molecule is
recorded at a constant applied force as a function of time as the
ssDNA is converted into dsDNA. From these measurements it is possible
to recover the number of replicated nucleotides as a function of time
and the speed of the motor. In all cases the replicating activity was
found to decrease with the applied force ceasing at around 40pN in the
case of the T7 DNApol above which exonuclease activity was
observed. Models of the mechanochemical cycle of DNApol have been
proposed for T7 which qualitatively reproduce the dependence of the
net replication rate as a function of the opposing load
\cite{GoeFraEllHer01,GoeDeaHer03}.  Another class of enzymes that move
in specific directions along dsDNA are translocases. These are
regulatory enzymes important during transport and replication
processes which sometimes show sequence dependent bidirectional
motion. FtsK, a translocase of {\em Escherichia coli} involved in
chromosome seggregation and cellular division, has been shown to move
at an average speed of 5 Kbp/s and against loads up to 60 pN
\cite{SalBigBarAll05,PeaLevCosGorPtaSheBusCoz05}.

\begin{figure}
\centering \includegraphics[scale=.5,angle=-90]{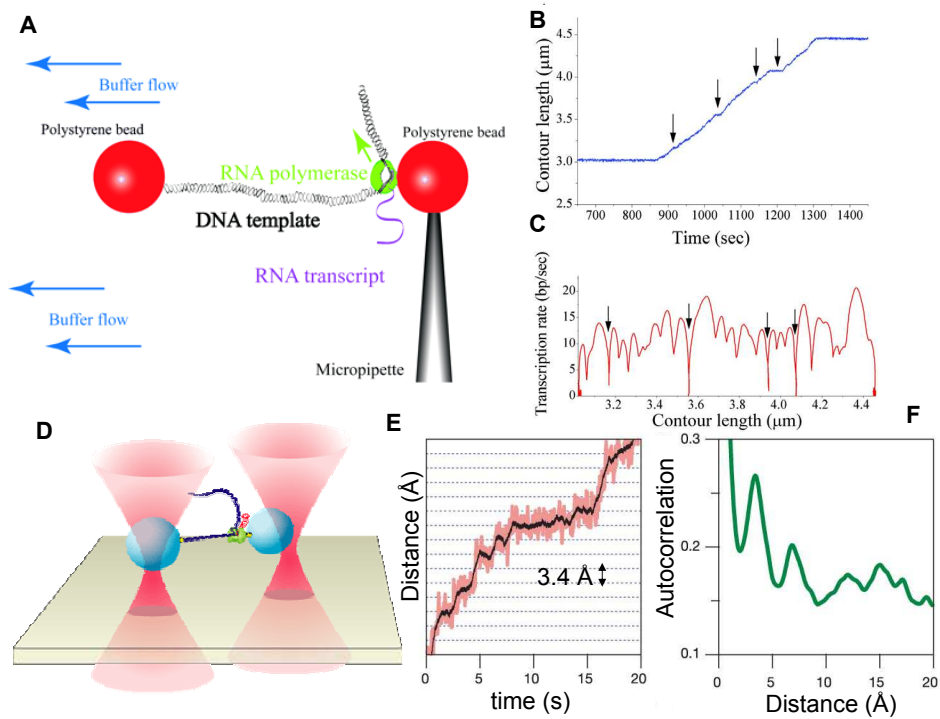}
  \caption{RNApol motion in {\em E. coli} using LOT. (A) Experimental setup in
  force-flow measurements. LOT are used to trap beads but forces
  are applied on the RNApol-DNA molecular complex using the Stokes drag force acting on the
  left bead immersed in the flow. In this scheme force assists RNA
  transcription as the DNA tether between beads increases in length as a
  function of time. (B) The contour length of the DNA tether as a
  function of time (blue curve) and (C) the transcription rate (red curve)
  as a function of the contour
  length. Pauses (temporary arrests of transcription) are shown as
  vertical arrows. (D) Experimental setup in a ultrastable LOT. The
  right trapped bead is located in the region of the trap where the
  stiffness vanishes. This creates a force-clamp where only the right
  bead moves determining the extension of the complex. (E) Motion of
  RNApol resolved in discrete steps of 3.4 \AA~. (F) Average
  autocorrelation function obtained from the position histograms showing
  peaks at distances multiples of 3.4 \AA~. Pictures (A-C) taken
  from \cite{ForIzhWooWuiBus02}. Pictures (D-F) taken from \cite{AbbGreShaLanBlo05}.}
\label{fig10}
\end{figure}

A related experimental setup has been applied to investigate the
transcription dynamics of RNApol. This is a processive enzyme that
synthesizes a ssRNA strand by translocating along dsDNA without the need
of a primer \cite{GelLan98}. The newly synthesized strand, messenger
RNA, codes for the amino acid sequence of proteins which are synthesized
in the ribosome \cite{Klug01}.  Transcription consists of three main
steps: initiation of the transcribing complex, elongation and
termination \cite{UptKanCham97}. Pioneering SME on transcription were
carried out in the T7 RNApol by the groups of Block, Gelles and Landick using LOT
\cite{YinWanSvoLanBloGel95,WanSchYinLanGelBlo98}. Under an applied force
the rate of RNA transcription did not change much but the process
stalled at around 20-30 pN. Similar results have been found in studies of
the transcription dynamics of {\em Escherichia coli} RNApol under
assisting or opposing force \cite{DavWuiLanBus00,ForIzhWooWuiBus02} (see
Fig.~\ref{fig10} (A-C)). The recent development of an ultra-stable trap
with Angstrom-level resolution has shown discrete steps of average
around 3.5 \AA~ for the step size of the enzyme indicative of a
mechanochemical cycle of one bp at a time
\cite{ShaAbbLanBlo03,AbbGreShaLanBlo05} (Fig.~\ref{fig10} (D-F)). The
dynamics of RNApol displays a complex behavior sensitive to DNA
sequence with pauses (temporary halts to transcription) and arrests
(permanent halts).  Many aspects of the elongation process in RNApol are
not well understood. For example the distribution of residence times for
pauses shows strong statistical variations in their frequency and
duration \cite{AdePorSanLisRobWan02}. Many of these features are also
found in DNA replication and are expected to be important in the
regulation of gene expression \cite{OrpRei02}.

Another type of DNA-protein motor that has attracted considerable
attention recently are virus packaging motors. The packaging motor of
bacteriophage virus $\phi 29$ encapsulates DNA inside the heads of
the virus. It has been shown to be a highly processive motor capable to
work against loads of up to approximately 50pN
\cite{SmiTanSmiGriAndBus01}. These experiments have stimulated very
interesting theoretical activity in the packaging
problem \cite{PurKonPhi03,KinTziBenGel01,ZanRegRudGel03}.

SME have been also applied to the study of topoisomerases. These are a
large class of enzymes present in both prokaryotes and eukaryotes and
are very important in the maintenance of the genome. Topoisomerases
act on the topology of DNA, their main task being to relax and
introduce supercoils in DNA by cutting the phosphate backbone of ssDNA
and dsDNA. For example, relaxing stress of supercoiled DNA is important for
transcription, replication and recombination. There are two major
families of topoisomerases: type I and type II. Type I topoisomerases
relax DNA supercoils by changing the linking number (for a definition
see Sec.~\ref{DNA}) in one unit
($\Delta L_k=-1$) without ATP consumption. Type II topoisomerases
change the linking number by 2 being capable of introducing supercoils
in DNA, $\Delta L_k=\pm 2$, therefore consuming ATP. SME using MT are
ideal to study the action of topoisomerases. By twisting the bead
attached to a DNA molecule it is possible to build up torsional stress
in the molecule and follow the subsequent relaxation upon addition of
topoisomerase in the buffer. Examples include the study of
topoisomerase II in {\em Drosophila melanogaster}
\cite{StrCroBen00,ChaBenCro03}, the observed torque dependence of
kinetics in eukaryotic topoisomerase I
\cite{DekRybDugCriCozBenCro02,KosCroDekShuDek05}. Also, DNA gyrase, a
topoisomerase II of prokaryotes known to introduce negative
supercoils, has been studied \cite{GorBryStoNolCozBus06} using the
bead tracking method \cite{BryStoGorSmiCozBus03}.  Related studies
have been conducted with {\em Escherichia coli} topoisomerase IV, an
ATP-dependent enzyme that removes positive (but not negative)
supercoils \cite{ChaStrBenCro05}. Most of these experiments have been
carried out using MT \cite{DesLioXiBenCro04} or SMF where limited
spatial resolution hinders time-dependent aspects of the kinetics such
as pauses or stalls, see however \cite{GorBryStoNolCozBus06}. 

Recent studies on helicases, however, have
already identified such effects. Helicases are yet another class of
DNA-protein motors crucial during DNA replication that have been
studied with SME.  Helicases are forerunners of the DNApol that unwind
DNA by transforming dsDNA into two strands of ssDNA, a necessary step
for the advance of the replication fork during DNA replication. DNA
unwinding has been studied in the RecBCD helicase/nuclease of {\em
E. coli} \cite{DohGel01,PerLiDalGelBlo04} and in the Rep helicase
using SMF assays \cite{HaRasCheBabGauLohChu02}. RNA helicases play
also an important part in the infection cycle of many viruses by
making two strands of ssRNA (available to produce new virus copies)
from a double stranded RNA (dsRNA) synthesized inside the infected cell. Recent studies of
the NS3 helicase, part of the protein machinery of hepatitis C virus,
has been recently studied using LOT showing a great richness of
kinetic effects including pauses, arrests or even reversal of the
motor followed by rewinding of the previously unwound dsRNA
\cite{DumCheSerBerTinPylBus06}. Finally we mention the study of the
molecular complex of the ribosome, an experimental challenge to the
biophysicist using single-molecule methods
\cite{VanVlaKnuGolCoo03,BlaKimGonPugChu04,VanTagShuCooGol05}.

Molecular motors have inspired a large number of theoretical studies
in statistical physics
\cite{Hill75,EisHil85,CorErmOst92,Astumian97,JulAdjPro98,FisKol99,KelBus00,LipKluNie01,Reimann02,BetJul05,Bhattacharjee04,KolStuPop05}. All
motors studied up to now show generic properties such as ATP and load
dependence of their average velocity.  For example the average speed
as a function of ATP concentration follows the Michaelis-Menten
kinetics. Also, there is some evidence that kinetic phenomena such as
pauses, arrests and backtracking motion are generic features of
motors. One wonders whether it exists a relationship between these
dynamical features of motors and their astonishing mechanical
efficiency.

\section{Tests of nonequilibrium theories in statistical physics}
\label{fts}
Recently there has been a lot of interest in applying single-molecule
techniques to explore physical theories in systems out of
equilibrium. The use of new micromanipulation tools in the exploration
of the behavior of tiny objects (such as biomolecules and motors)
embedded in a thermal environment opens the possibility to investigate
how these systems exchange energy with their environment. This question
is of great interest both at a fundamental and practical level.  From a
fundamental point of view, the comprehension of how biomolecules
operating very far from equilibrium are so efficient (see discussion in
Sec.~\ref{physbioback}) raises the question whether such tiny systems
exploit rare and large deviations from their average behavior by
rectifying thermal fluctuations from the bath. From a practical point of
view, this might help in the design of efficient nanomotors in the
future. The study of such questions is steadily becoming an active area
of research, nowadays referred to as {\em Nonequilibrium thermodynamics
of small systems} \cite{BusLipRit05,Ritort06}. Such discipline is becoming quite
popular among statistical physicists who recognize there new aspects of
thermodynamics where large Brownian fluctuations are of pivotal
importance as compared to fluctuations in macroscopic (or large) systems
\cite{FreKro05}. In macroscopic systems, fluctuations represent just
small deviations respect to the average behavior. For example, an ideal
gas of $N$ molecules in thermal contact with a bath at temperature $T$ has
an average total kinetic energy of $(3/2)Nk_BT$. However, the total
energy is not a conserved quantity but fluctuates, its spectrum being a
Gaussian distribution of variance $(3/2)N(k_B T)^2$. Therefore, relative
deviations of the energy are on the order $1/\sqrt{N}$ respect to the
average value. For macroscopic systems such deviations are very small:
for $N=10^{12}$ (this is the typical number of molecules in a 1 ml test
tube at nanomolar concentrations), then relative deviations are on the
order of $10^{-6}$, hence experimentally unobservable by calorimetry
methods. For a few molecules, $N\sim {\cal O}(1)$ relative deviations
are on the same order and fluctuations are measurable.  SME, by allowing
to study molecules one at a time, grant access to such large deviations
that are inaccessible in bulk experiments which use a macroscopic number
of molecules. As a rule of thumb we can say that in nonequilibrium
processes in small systems the typical amount of energy exchanged with
the environment is a few times $k_BT$, maybe from 1 to 1000 but not much
more. As often happens when establishing the limits of validity of
certain regimes, there is not a well defined frontier separating the
small-size regime from the large-size regime.

The name thermodynamics of small systems was already coined by
T. L. Hill \cite{Hill94} who showed the importance of the statistical
ensemble in thermodynamic relations. A main result of statistical
mechanics is the independence of the equation of state on the
statistical ensemble in the thermodynamic limit. Such independence
breaks down in small systems due to the contribution of fluctuations
which depend on the type of statistical ensemble considered (e.g for the
case of a stretched polymer \cite{KelSwiBus03}). The search for a new
thermodynamic description of small systems has given rise to
microcanonical ensemble theory of phase transitions \cite{Gross01} and
new classical statistics such as that embodied in Tsallis-entropy
\cite{Tsallis88} and Beck's theory \cite{Beck01} (for a review see
\cite{Cohen02}). From the current point of view, the most important
aspect of biomolecular complexes is that they operate far from
equilibrium, yet the possible relationship between nonequilibrium
behavior and biological function is still unknown.  The combination of small size and
nonequilibrium behavior appears as the playground for the striking behavior
observed in molecular complexes inside the living cell.

Since the beginning of the 90's some exact results in statistical
mechanics have provided a mathematical description of energy
fluctuations (in the form of heat and work) for nonequilibrium systems.
This new class of results go under the name of fluctuation theorems
(FTs) and provide a solid theoretical basis to quantify energy
fluctuations in nonequilibrium systems \cite{EvaSea02,Maes03,Ritort06}. FTs
describe energy fluctuations in systems while they execute transitions
between different types of states. For these fluctuations to be
observable the system has to be small enough and/or operate over short
periods of time, otherwise the measured properties approach the
macroscopic limit where fluctuations get masked by the dominant average
behavior. Most fluctuation theorems are of the form,
\be
\frac{P(+{\cal S})}{P(-{\cal S})}=\exp\Bigl(\frac{{\cal S}}{k_B}\Bigr)~~~~,
\label{FT}
\ee
where ${\cal S}$ has the dimensions of an entropy that may represent
heat and/or work produced during a given time interval. The precise
mathematical form of relations such as \eq{FT} (for instance, the
precise definition of ${\cal S}$ or whether they are valid at finite time
intervals or just in the limit where the time interval goes to
infinity) depends on the particular nonequilibrium conditions
(e.g. whether the systems starts in an equilibrium Gibbs state, or
whether the system is in a nonequilibrium steady state, or whether the system executes
transitions between steady states, etc..).

Various categories of FTs have been introduced and experimentally
validated \cite{Ritort06}. The differences between various FTs can be illustrated by
introducing the concept of the control parameter. The control parameter
(let us say $\lambda$) is a value or a set of values that, once
specified, fully characterizes a given stationary state of the system
(either equilibrium or nonequilibrium). Upon variation of the control
parameter a system that is initially in a well defined state will evolve
toward a new state. In general, if the control parameter is varied with
time according to a given protocol, $\lbrace \lambda(t); 0\le t \le t_f\rbrace$
the system will evolve along a given trajectory or path. If, after time
$t_f$, the value of the control parameter is kept fixed at the value
$\lambda(t_f)$ then the system will eventually settle into a new
stationary state. Along a given path the system will exchange energy
with its environment in the form of heat and work. The values of the
heat and work will depend on the path followed by the system. Upon
repetition of the same experiment an infinite number of times (the
protocol $\lambda(t)$ being the same for all experiments), there will be
a heat/work distribution characterizing the protocol
$\lambda(t)$. Generally speaking, FTs relate the amounts of work or heat
exchanged between the system and its environment for a given
nonequilibrium process and its corresponding time-reversed process.  The
time-reversed process is defined as follows. Let us consider a given
nonequilibrium process (we call it forward, denoted by F) characterized
by the protocol $\lambda_F(t)$ of duration $t_f$. In the reverse process
(denoted by R) the system starts at $t=0$ in a stationary state at the
value $\lambda_F(t_f)$ and the control parameter is varied for the same
duration $t_f$ as in the forward process according to the protocol
$\lambda_R(t)=\lambda_F(t_f-t)$.  FTs depend on the type of initial
state and the particular type of dynamics (deterministic versus
stochastic) or thermostatted conditions.  

Despite of the fact that most of these theorems are treated as distinct they are in fact closely
related \cite{Ritort06}. The main hypothesis for all theorems is the
validity of some form of microscopic reversibility or local detailed
balance (see however \cite{CohMau04,Jarzynski04,Astumian05}). Major
classes of FTs include the transient FT (TFT) and the steady state FT (SSFT):
\begin{itemize}

\item{The transient FT (TFT).} In the TFT the system initially starts in an equilibrium
(Boltzmann-Gibbs) state and is driven away from equilibrium by the
action of time-dependent forces that derive from a time-dependent potential
$V_{\lambda(t)}$. The potential depends on time through the value of the control
parameter $\lambda(t)$. At any time during the process the system in an
unknown transient nonequilibrium state. If the value of $\lambda$ is
kept fixed then the system relaxes into a new equilibrium state.
The TFT was introduced by Evans and Searles \cite{EvaSea94} in
thermostatted systems and later extended by Crooks to Markov processes
\cite{Crooks99}.

\item{The steady state FT (SSFT).} In the SSFT the system is in a
nonequilibrium steady state where it exchanges net heat and work with the
environment. The existence of the SSFT was numerically anticipated by 
Evans and collaborators for thermostated systems  \cite{EvaCohMor93} and
demonstrated for deterministic Anosov systems by Gallavotti and Cohen
\cite{GalCoh95}. The entropy production ${\cal S}$  by the system (equal to the heat
exchanged by the system divided by the temperature of the environment) satisfies the relation
\eq{FT} in the asymptotic limit of large times $t\to \infty$ and for
bounded energy fluctuations, $\sigma=\frac{|{\cal S}|}{t}<\sigma^*$
where $\sigma^*$ is a model dependent quantity. 
Other class of SSFTs include stochastic dynamics \cite{Kurchan98}, Markov
chains \cite{LebSpo99,Maes99} or the case where the system initially starts in a
steady state \cite{OonPan98} and executes transitions between different
steady states \cite{HatSas01,SpeSei05}.

\end{itemize}

The first experimental tests of FTs were carried out by Ciliberto and coworkers
for the Gallavoti-Cohen FT in Rayleigh-Bernard convection
\cite{CilLar98} and turbulent flows \cite{CilGarHerLacPinRui04}.  Later
on FTs were tested for beads trapped in an optical potential and
moved thorugh water at low Reynolds numbers. The motion of the bead is then
well described by a Langevin equation that includes a friction
(non-conservative) force, a confining (conservative) potential and a
source of stochastic noise. Experiments have been carried out by Evans
and collaborators who have tested the validity of the TFT
\cite{WanSevMitSeaEva02,CarReiWanSevSeaEva04}, and by Liphardt and
collaborators for a bead executing transitions between different steady
states \cite{TreJarRitCroBusLip04}. The validity of the TFT has been
also recently tested for non-Gaussian optical trap potentials
\cite{BliSpeHelSeiBech05}.

Particularly relevant to the single molecule context is the FT by Crooks
\cite{Crooks99,Crooks00} which relates the work
distributions measured along the forward (F) and reverse (R) paths,
\be
\frac{P_F(W)}{P_R(-W)}=\exp\Bigl( \frac{W-\Delta G}{k_BT}\Bigr)~~~~~,
\label{CrooksFT}
\ee
where $P_F(W),P_{R}(-W)$ are the work distributions along the F and R
processes respectively, and $\Delta G$ is the free energy difference
between the equilibrium states corresponding to the final value of the
control parameter $\lambda_f=\lambda(t_f)$ and the initial one
$\lambda_i=\lambda(0)$. A particular result of \eq{CrooksFT} is the
Jarzynski equality \cite{Jarzynski97,Jarzynski02} that is obtained from
\eq{CrooksFT} by rewriting it as $P_R(-W)=\exp\bigl( \frac{-W+\Delta
G}{k_BT}\bigr)P_F(W)$ and integrating both sides of the equation
between $W=-\infty$ and $W=\infty$. Because of the normalization
property of probability distributions, the left hand side is equal to 1
and the Jarzynski equality reads,
\bea
<\exp\Bigl( -\frac{W}{k_BT}\Bigr)>_F=\exp\Bigl( -\frac{\Delta
G}{k_BT}\Bigr)~~{\rm or}~~~\nonumber\\\Delta G=-k_BT\log\Bigl(<\exp\bigl( -\frac{W}{k_BT}\bigr)>_F\Bigr)~~~~,
\eea
where $<...>_F$ denotes an average over an infinite number of
trajectories all them generated by a given forward protocol
$\lambda_F(t)$. Relations similar to Jarzynski's equality can be traced
back in the free-energy perturbation identity derived by Zwanzig
\cite{Zwanzig54} and a generalized fluctuation-dissipation relation
proposed by Bochkov and Kuzovlev \cite{BocKuz81}.  The Jarzynski
equality and the FT by Crooks can be used to recover equilibrium
free-energy differences between different molecular states by using
nonequilibrium SME using LOT \cite{HumSza00,Jarzynski01,HumSza05}. In
2002, the Jarzynski equality was tested to pull the P5ab RNA hairpin, a
derivative of the {\em Tetrahymena Termophila} L21 ribozyme
\cite{LipDumSmiTinBus02}. However, in that case the molecule was pulled
not too far from equilibrium. The Jarzynski equality and related
identities for athermal systems have been recently put under scrutiny in
other systems \cite{SchSpeTieWraSei05,Douarche05a,Douarche05b}. The
Jarzynski equality and the FT by Crooks have inspired several
theoretical papers discussing other related exact results
\cite{ZonCoh03a,ZonCoh03b,Seifert04a,JarWoj04,Seifert05a,ReiSevEva05},
free-energy recovery from numerical simulations
\cite{Hummer01,IsrGaoSch01,JenParTajSch02,ParKhaTajSch03,AndDinKar03,ParSch04},
bias and error estimates for free-energy differences
\cite{Bennett76,WooMuhTho91,HenJar01,ZucWol02a,ZucWol02b,GorRitBus03,ShiBaiHooPan03},
applications either to single-molecule pulling experiments
\cite{RitBusTin02,SchFuj03,BraHanSei04}, enzyme kinetics
\cite{Qian05,MinJiaYuKouQiaXie05} or solvable models
\cite{MazJar99,LuaGros05,BenBroKaw05,Seifert05b,SpeSei05b,CleBroKaw06}. In
addition, analytical studies on small systems thermodynamics show that
work/heat distributions display non-Gaussian tails describing large and
rare deviations from the average and/or most probable behavior
\cite{ZonCoh03a,ZonCoh03b,ZonCoh04,Ritort04,ImpPel05b,ImpPel05c}. These
theoretical studies open the way to investigate the possible relevance
of these large deviations in other nonequilibrium systems in condensed
matter physics \cite{CilLar98,CilGarHerLacPinRui04,GarCil04,FeiMen04}.

\begin{figure}
\centering \includegraphics[scale=.55,angle=-90]{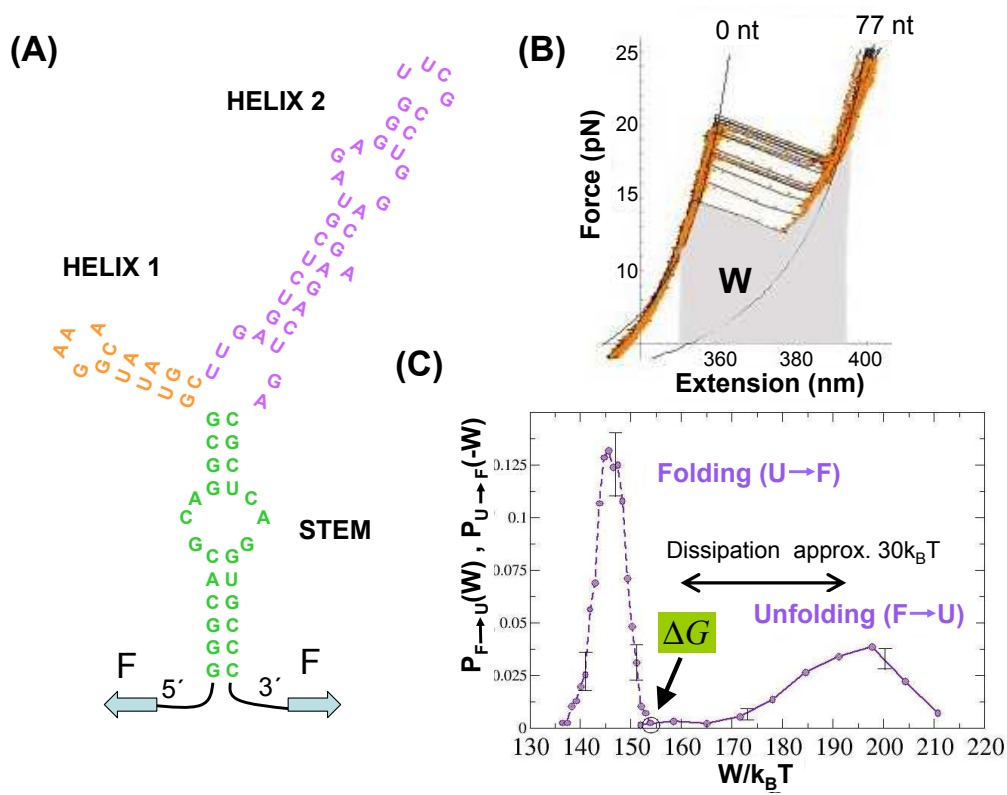}
  \caption{Recovery of folding free energies in a three-helix junction
  RNA molecule \cite{ColRitJarSmiTinBus05}. (A) Secondary structure of
  the junction containing one stem and two helices. (B) Typical
  force-extension curves during the unfolding process. The gray area
  corresponds to the work exerted on the molecule for one of the
  unfolding curves. (C) Work distributions for the unfolding or forward
  paths ($F\to U$) and the refolding or reverse ($U\to F$) paths
  obtained from 1200 pulls. According to the FT by Crooks \eq{CrooksFT}
  both distributions cross at $W=\Delta G$. After subtracting the free
  energy contribution coming from stretching the handles and the ssRNA
  these measurements provide a direct measure of the free energy of the
  native structure. }
\label{fig12}
\end{figure}

The FT by Crooks can be applied and tested by measuring the unfolding and refolding
work distributions in single molecule pulling experiments. For example,
let us consider the case of a molecule (e.g. a DNA or RNA hairpin or a
protein) initially in thermal equilibrium in the folded (F) or native
state. By applying mechanical force (e.g. using AFM or LOT) the molecule can be mechanically
unfolded and the conformation of the molecule changed from the native to
the unfolded (U) state.  The unfolding event is
observed by the presence of a rip in the FEC of the molecule
(Fig.~\ref{fig12}B). During the unfolding process the tip of the
cantilever or the bead in the trap exerts a mechanical work on the
molecule that is given by,
\be
W=\int_{x_0}^{x_f} Fdx
\label{work}
\ee
where $x_0,x_f$ are the initial and final extension of the molecule. In
\eq{work} we are assuming that the molecular extension $x$ is the
externally controlled parameter (i.e. $\lambda\equiv x$) which is not
necessarily the case. However the corrections introduced by using
\eq{work} are shown to be often small. The work \eq{work} done upon the
molecule along a given path corresponds to the area below the FEC that
is limited by the initial and final extensions, $x_0$ and $x_f$ (grey shaded
area in Fig.~\ref{fig12}B). Because the unfolding of the
molecule is a stochastic (i.e. random) process, the value of the force
at which the molecule unfolds changes from experiment to experiment and
so does the value of the mechanical work required to unfold the
molecule. Upon repetition of the experiment many times a distribution of
unfolding work values for the molecule to go from the folded (F) to the
unfolded (U) state is obtained, $P_{F\to U}(W)$. A related work
distribution can be obtained if we reverse the pulling process by
releasing the molecular extension at the same speed at which the
molecule was previously pulled, to allow the molecule to go from the
unfolded (U) to the folded (F) state. In that case the molecule refolds
by performing mechanical work on the cantilever or the optical
trap. Upon repetition of the folding process many times the work
distribution, $P_{U\to F}(W)$ can be also measured. The unfolding and
refolding work distributions can then be measured in
stretching/releasing cycles, an example is shown in Fig.~\ref{fig12}C.

The FT by Crooks has been tested in different types of RNA
molecules and the method has been shown capable of recovering
free-energies under strong nonequilibrium conditions
\cite{ColRitJarSmiTinBus05}. From \eq{CrooksFT} we observe that
$P_F(\Delta G)=P_R(-\Delta G)$ so the forward and reverse work
probability distributions cross each other at $W=\Delta G$. By
repeatedly measuring the irreversible mechanical work exerted upon the
molecule during the unfolding process and the mechanical work
delivered by the molecule to the LOT instrument during the refolding
process it has been possible to reconstruct the unfolding and refolding
work probability distributions, $P_{F\to U}(W)$ and $P_{U\to F}(-W)$, and extract the
value of the work $W=\Delta G$ at which both distributions cross each
other (Fig.~\ref{fig12}C).  The work probability distributions where measured along the
unfolding and refolding pathways for a three-way junction RNA molecule
and found to strongly deviate from a Gaussian distribution
\cite{ColRitJarSmiTinBus05} (Fig.~\ref{fig12}C). These experimental results pave the way
for other related studies, for example in molecular dynamics
simulations \cite{KosBarJan05}.

These kind of studies will expand in the future to cover more
complex cases and other nonequilibrium situations such as the
free-energy recovery of folding free energies in native states in
proteins or free energies in misfolded structures and intermediate
states in RNA molecules and proteins. Ultimately FTs, when combined with SME,
will offer an excellent opportunity to characterize and understand the
possible biological relevance of large deviations and extremal
fluctuations in molecular systems.

\section{Conclusions}
\label{conclusions}

In this review I have discussed the potential of SME to investigate
various topics in molecular biophysics and statistical mechanics. After
a brief discussion of the most widely used experimental techniques I
have presented applications to various molecular systems. As stressed in
the Introduction this review does not exhaust all relevant applications
of SME. By focusing on the field of molecular biophysics I just covered a
small fraction of problems. Other areas such as cellular biophysics and
condensed matter physics are progressively incorporating such techniques in
the labs.

What is the future of SME? We can foresee two aspects of SME whose
development will be crucial for the progress in the field:
development of new and better instruments and development of new
and better protocols of chemical synthesis. 

A major contribution to the progress of the field will come from
instrumentation design with enhanced spatial and temporal resolution,
Recently, the development of an ultra-stable optical trap with
Angstrom-level resolution has allowed for the direct observation of
base-pair stepping during the transcriptional elongation of {\em
E. coli} RNApol \cite{ShaAbbLanBlo03,AbbGreShaLanBlo05}.  Future developments to 
improve spatial detection will certainly include optical tweezers with dual traps \cite{MofCheIzhBus06}. Combination
of SMF techniques for imaging with force measurements open also the
way to more sensitive measurements.  Single molecule FRET can be
combined with LOT to measure forces and correlate them with
conformational changes \cite{LanForBlo03}.  Also total internal
reflection fluorescence (TIRF) techniques capable of monitoring the position
of a molecule along a vertical direction using a calibrated evanescent
wave can be combined with AFM measurements \cite{SarRobFer04}.
Instruments that can manipulate molecules at different temperatures
and forces will grant access to the potential energy surface of the
molecule. Along this direction, AFM and LOT that include various temperature
controllers have been already developed
\cite{WilWenRouBlo01,LawLiaYanSpeDis03,MaoAriSmiTinBus05}.  Finally, LOT with multiple
traps offer interesting possibilities in the near future although it
is not yet clear how to use them to manipulate molecular complexes in
a controlled way \cite{GarGloMelSibDho02}.  Holographic tweezers offer
also exciting possibilities \cite{Grier03} however we must first learn
how to calibrate them in order to measure forces.

Also, the development of better protocols to synthesize molecular
systems will facilitate the outcome of the experiments. SME may
present a considerable difficulty regarding the preparation of the
samples, specially in those cases where biological activity is
required. It is commonly said that SME are $100\%$ noise and $100\%$
signal. Uncertainties in experimental conditions and sample
preparation imply that experiments have to be repeated several times
until good results are obtained. Usually, just a few molecules show
the interesting behavior one is looking for, the rest simply do not
work.  By synthesizing better complexes and having a good control on
the chemistry it will be possible to carry out more efficient
experiments and investigate new problems. A good example of
potentially interesting SME, where the chemical synthesis of the
molecular complex to manipulate is the rate-limiting step, is found in
the study of intermolecular interactions, for example protein-protein
interactions (Sec.\ref{proteinprotein}).  Most studies on
intermolecular interactions use the AFM. Unfortunately with such
technique it is difficult to repeteadly form and dissociate the same
set of molecular interactions. This experiment is more feasible
using LOT where, in addition, non specific interactions between
substrate and the molecule are more efficiently avoided. The chemical
synthesis of appropriate handles (e.g. carbon nanotubes) for protein
complexes will facilitate such experiments. We can foresee future SME
to investigate protein-protein aggregation processes that are crucial
in many biological processes.

SME are fascinating but difficult at the same time. They require
imagination and strenuous efforts. SME are redefining the shape and
composition of modern research teams. These must include researchers
with expertises and knowledge covering a wide range of topics. SME
represent one of the most interdisciplinary fields at present that will
require the combined efforts of biologists, chemists and physicists. SME
represent a good example of the new trends in modern science that will
reshape the way we are going to do research in this century.

{\bf Acknowledgments}.  I warmly thank M. Carrion-Vazquez for
useful suggestions on the manuscript. Observations by an anonymous
referee are also acknowledged. I wish to thank also S. Dummont,
J. Gore, M. Manosas and P. Sollich for a critical reading of parts of the manuscript. I
acknowledge the warm hospitality during past years of the Bustamante,
Tinoco and Liphardt labs at UC Berkeley where I had the great
opportunity to learn about this fascinating subject.  I am grateful to
all my collaborators, much of whose work has been described in this
review, including C. Bustamante, D. Collin, G. E. Crooks, J. Gore,
J. M. Huguet, C. Jarzynski, I. Junier, P. Li, J. Liphardt, M. Manosas,
S. Mihardja, D. Navajas, S. Saxon, S. Smith, R. Sunyer, I. Tinoco, E. Trepagnier,
J. D. Wen. This work has been supported by the European community
(STIPCO network), the SPHINX ESF program, the Spanish research council
(Grants FIS2004-3454,NAN2004-09348) and the Catalan government
(Distinci\'o de la Generalitat 2001-2005).

\section{List of abbreviations}
\begin{description}

\item[ADP] Adenosine diphosphate
\item[AFM] Atomic force microscopy 
\item[ATP] Adenosine triphosphate
\item[BFP] Biomembrane force probe 
\item[dsDNA] Double stranded DNA
\item[dsRNA] Double stranded RNA
\item[DNA] Deoxyribonucleic acid
\item[DNApol] DNA polymerase 
\item[FEC] Force-extension curve
\item[FRET] Fluorescence resonance energy transfer
\item[FT] Fluctuation Theorem
\item[MT] Magnetic tweezers
\item[LOT] Laser optical tweezers
\item[RNA] Ribonucleic acid
\item[RNApol] RNA polymerase 
\item[SME] Single-molecule experiments
\item[SMF] Single-molecule fluorescence
\item[ssDNA] Single stranded DNA
\item[ssRNA] Single stranded RNA
\end{description}

\section*{References}
\bibliographystyle{unsrt}
\bibliography{reviewSM.rev4condmat}

\end{document}